# Alignment of metabolic trajectories with application to metabonomic toxicology

Dr Mansour Taghavi Azar Sharabiani

Imperial College London, 2005

# Table of contents:

# Abstract:

Geometry of the metabolic trajectories is characteristic of the biological response (Keun, Ebbels et al. 2004). Yet, due to unavoidable inter-individual variations, the exact trajectories characterising the biological responses differ. We examined whether the differences seen between metabolic trajectories of a specific treatment, correspond to the variations seen in the other biological manifestations of the same treatment. Differences in trajectories were measured via alignment procedures which introduced and implemented in this study. Our study revealed strong correlation between the scales of the aligned trajectories of metabolic responses and the severity of the hepatocelluar lesions induced after administration of hydrazine. Thus the results confirm that aligned trajectories are characteristic of a specific treatment. They then can be used for comparison with other treatment specific or unknown metabolic trajectories and can have many metabonomic applications such as preclinical toxicological screening

# 1. Introduction

High expectations of rapid drug discovery in post-genomics era via mining the wealth of the genome data proved to be rather optimistic due to the complexity of disease biology and the fact that identifying a target does not necessarily reveal what the target does or how it works. Research and development cost of new drugs continues to rise above general price inflation while approval rates fall (DiMasi, Hansen et al. 2003). These rather disappointing facts pronounce the importance of better understanding of biological responses to various signals not only based on relying on transcriptomic information, but also considering much broader perspective encompassing proteomics (Blackstock and Weir 1999) and some metabonomics(Nicholson, Lindon et al. 1999) data increase efficiency of drug discovery and development efforts (Butcher, Berg et al. 2004). Figure 1, extracted from a published paper (Lindon, Holmes et al. 2003), schematically demonstrate relationships between the genome and the technologies for evaluating changes in gene expression (transcriptomics), protein levels (proteomics), and small molecule metabolite effects (metabonomics). Some of the main categories of resources available to develop an ideal system capable of pre-clinically prediction of compounds having toxicity and unfavourable pharmacokinetic profiles before are structure–activity relationships (SAR) computational models based on compound structure; 'pattern' databases of tissue or organ response to drugs, compiled from high-throughput experiments such as COMET (Lindon, Keun et al. 2005) project; and 'systems biology' databases of metabolic pathways, genes and regulatory networks (Bugrim, Nikolskaya et al. 2004). Metabolite profiling has already been used throughout clinical development for discovery of clinical safety biomarkers (Nicholson, Connelly et al. 2002).

Metabonomic expert systems capable of predicting the probability of novel drug candidates being toxic employ series of mathematical models can be constructed operating at three levels i.e. classification of 'normal' samples, classification of samples of known toxicity or disease as described within a pre-existing database, and identification of biomarkers of pathology (Parzen 1962; Holmes, Nicholson et al. 2001; Holmes and Antti 2002). These models may include e.g. neural network(Parzen 1962)-based methods (D. S. Broomhead 1988; Bishop 1995), non-linear approaches such as probabilistic neural networks (PNN)(Specht 1990), non-neural implementation of probability classification technique (Ebbels, Keun et al. 2003) and/or PCA-based methodologies.

## 1.1. The 'Omics' approach

The 'Omics' interrelationship can be symbolised by a study (Sharabiani, Siermala et al. 2005), which shows during cellular responses and in many biological systems, gene expression strongly correlates with amino acid composition, physicochemical, and structural properties of proteins as well as the functional, cellular localization and gene ontology parameters of proteome. Figure 1, extracted from a published paper (Lindon, Holmes et al. 2003), schematically demonstrate relationships between the genome and the technologies for evaluating changes in gene expression (transcriptomics), protein levels (proteomics), and small molecule metabolite effects (metabonomics).

Omics-integration (i.e. genomics, proteomics and metabonomics) can characterise the response of living organisms to chemical exposure or other stimuli in terms of gene and protein expression as well as metabolic regulation and provide mechanistic and somewhat quantitative information which can be used to incorporate toxicological data at earlier stages of drug development, potentially saving millions of dollars (Burchiel, Knall et al. 2001; Aardema and MacGregor 2002; Keun, Ebbels et al. 2002; Tennant 2002).

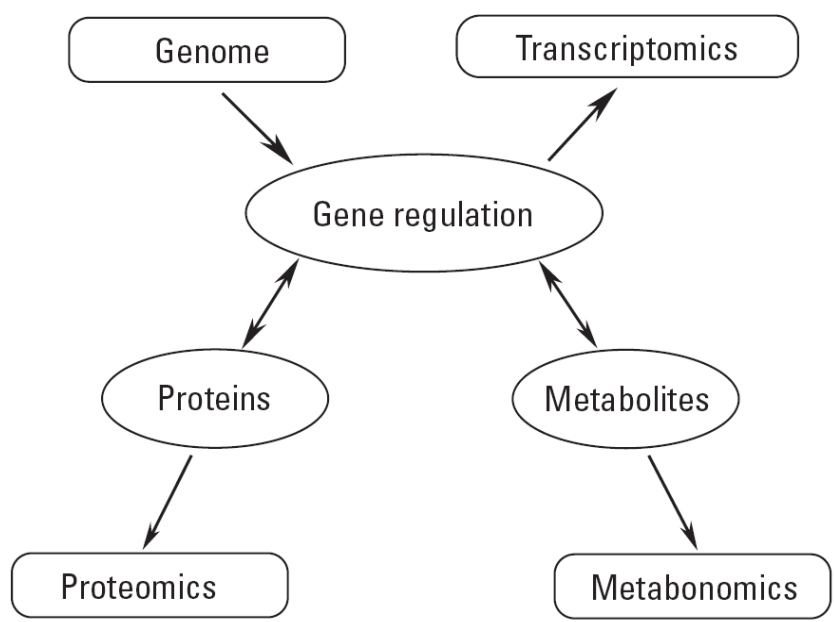

**Figure 1: (Extracted from a published paper(Lindon, Holmes et al. 2003)) Relationships between the genome and the technologies for evaluating changes in gene expression (transcriptomics), protein levels (proteomics), and small molecule metabolite effects (metabonomics)**

Omics techniques have been adopted by pharmaceutical industry to complement the traditional approaches of target identification and validation, and experimental analysis of traditional hypothesis-based methods (Butcher, Berg et al. 2004). Metabonomics can provide information in a non-invasive or minimally invasive manner through biofluids, such as blood plasma, urine, or cerebrospinal fluid (Lindon, Holmes et al. 2003) . Compared to the rest of omics technologies, which measure the intermediate steps in the response of the organism to the xenobiotics, has the important advantage of dealing with the 'end points' of protein action (Lindon, Holmes et al. 2003; Bugrim, Nikolskaya et al. 2004).

## 1.2. *Metabonomics & Metabolomics*

Although the terminology is still evolving, metabonomics has been defined as "quantitative measurement of time-related multi-parametric metabolic responses of multi-cellular systems to pathophysiological stimuli or genetic modification" (Nicholson, Lindon et al. 1999) or "measure of the fingerprint of biochemical perturbations cause by disease, drug or toxins"(Goodacre, Vaidyanathan et al. 2004) or it can be viewed as "the process of defining multivariate metabolic trajectories that describe the systemic response of organisms to physiological perturbations through time"(Keun, Ebbels et al. 2004). Metabonomics, an application-deriven science, has its roots in metabolite profiling (analysis focused on a group of metabolites e.g. those associated with a specific pathway (Fiehn 2002)) initially appeared in literature in 1950s (Rochfort 2005),. Metabolite profiling in a wider range is called metabolomics, defined as comprehensive analysis of the whole metabolome under a given set of conditions (Fiehn 2002; Goodacre, Vaidyanathan et al. 2004). The technologies commonly exploited for different metabolomic strategies are shown in Figure 2 extracted from a published paper (Goodacre, Vaidyanathan et al. 2004).

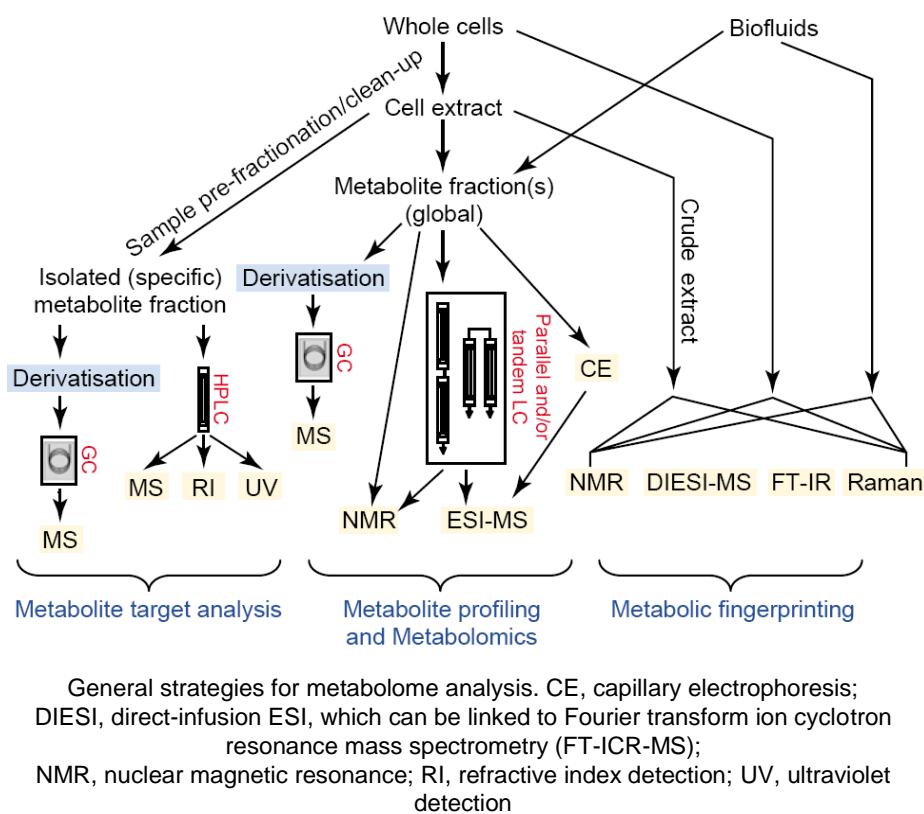

General strategies for metabolome analysis. CE, capillary electrophoresis; DIESI, direct-infusion ESI, which can be linked to Fourier transform ion cyclotron resonance mass spectrometry (FT-ICR-MS); NMR, nuclear magnetic resonance; RI, refractive index detection; UV, ultraviolet detection

**Figure 2: (Extracted from a published paper(Goodacre, Vaidyanathan et al. 2004)) Technologies for metabolome analysis**

NMR as well as Mass Spectroscopy (MS) is useful tools in metabonomics. NMR is non-destructive and cheap. With no major sample preparation, biofluids can be analysed by nuclear magnetic resonance (NMR), which can provide valuable information on metabolites (Lindon, Holmes et al. 2003; Nicholson and Wilson 2003).

The NMR-based metabonomic data have presented a high degree of stability without compromising the diagnostic power to identify the source and effects of confounding variation (Keun, Ebbels et al. 2002). MS is more sensitive than NMR and therefore can detect much smaller concentrations of metabolites. MAS (magic angle spinning) NMR(Andrew 1984) can be used in intact tissues for instance to correlate patterns of metabolic perturbation in the tissues with changes in the biofluid profile which often gives powerful insights such as defining diagnostic set of non-invasive biomarkers of toxicity and characterising the biochemical composition of healthy and diseased tissues (Cheng, Lean et al. 1996; Cheng, Ma et al. 1997; Millis, Maas et al. 1997; Moka, Vorreuther et al. 1997; Tomlins, Foxall et al. 1998; Garrod, Humpfer et al. 1999; Bollard, Garrod et al. 2000; Waters, Garrod et al. 2000; Garrod, Humpher et al. 2001).

Metabonomics recently has become a major focus of research in various areas (Kanehisa, Goto et al. 2002) especially indications of drug toxicity which is particularly addressed (Ellis, Broadhurst et al. 2002; Famili, Forster et al. 2003; Forster, Famili et al. 2003; Wilson and Nicholson 2003; Rochfort 2005). The COMET (Consortium for Metabonomic Toxicology) project is probably the biggest coordinated effort in metabonomics aimed at studying xenobiotic toxicity rigorously and comprehensively using mainly $^1$H NMR spectroscopy to classify the biofluids in terms of known pathological effects caused by administration of substances causing toxic effects (Lindon, Nicholson et al. 2003)

Five major pharmaceutical companies namely Bristol-Myers-Squibb, Eli Lilly & Co., Hoffman-La Roche, NovoNordisk, Pfizer Inc. and The Pharmacia Corporation (now within Pfizer) and Imperial College London, UK took part in the consortium aiming at generating a metabonomic database of the effects of 147 model toxins and other treatments. ~500 000 $^1$H nuclear magnetic resonance (NMR) spectra of rodent urine and blood serum and developing a predictive expert system for target organ toxicity (Lindon, Nicholson et al. 2003; Lindon, Nicholson et al. 2003; Lindon, Keun et al. 2005).

## 1.3. Statistical analysis & Expert systems in toxicological screening:

Multivariate statistical methods were initially introduced to metabolomics to analyse to simply to classify objects (e.g. 1H NMR spectra of biofluid samples) according to their origin (i.e. the site or mechanism of toxic exposure). (Gartland, Beddell et al. 1991; Holmes and Antti 2002). As metabolomics applications and objectives evolve, the range of statistical methodologies being introduced to metabonomics increases and diversifies. A wide range of PCA based methods such as partial least squares (PLS) or PLS-Discriminant Analysis (PLS-DA), batch processing (BP), soft independent modelling of class analogy (SIMCA) have been developed and employed in metabonomic studies. Other method also include nonlinear mapping procedures (NLM), hierarchical cluster analysis (HCA), probabilistic models, probabilistic neural networks (PNN), as well as non-neural implementation of probability classification technique, various scaling techniques, geometric classification analysis, ANOVA-simultaneous component analysis (Nicholson, Buckingham et al. 1983; Gartland, Beddell et al. 1991; Anthony, Sweatman et al. 1994; Nicholson, Foxall et al. 1995; Robertson, Reily et al. 2000; Ebbels, Lindon et al. 2001; Holmes, Nicholson et al. 2001; Lindon, Holmes et al. 2001; Nicholls, Holmes et al. 2001; Waters, Holmes et al. 2001; Waters, Holmes et al. 2001; Antti, Bollard et al. 2002; Azmi, Griffin et al. 2002; Brindle, Antti et al. 2002; Ebbels 2002; Ellis, Broadhurst et al. 2002; Holmes and Antti 2002; Keun, Ebbels et al. 2002; Beckonert, Bollard et al. 2003; Ebbels, Keun et al. 2003; Keun, Ebbels et al. 2003; Lenz, Bright et al. 2003; Wang, Bollard et al. 2003; Antti, Ebbels et al. 2004; Bugrim, Nikolskaya et al. 2004; Ebbels, Holmes et al. 2004; Keun, Ebbels et al. 2004; Kleno, Kiehr et al. 2004; Bollard, Keun et al. 2005; Dyrby, Baunsgaard et al. 2005; Smilde, Jansen et al. 2005; Hector C. Keun 2006; Timothy M. D. Ebbels 2006). While unsupervised methods such as PCA are useful for comparing normal and test samples, with large number of classes, supervised methods may be advantageous owing discrimination between classes. Partial least-squares (PLS), a supervised method, which by associating independent variables to dependent variables can reveal the influence of time or other external variables, particularly useful for analysis of samples taken over a period of time. Discriminant analysis is used to establish the optimal position to place a discriminant surface that best separates classes. SIMCA, a supervised method classifies data and models formed based on the training set. PCA is performed for each class (training set) and each sample in the test set is assigned into class model which can be best predicted. Therefore, SIMCA has the flexibility of not assigning samples into an inappropriate category if no class can be assigned. Similarly, following preliminary unsupervised analysis such as PCA, the supervised Partial Least-Squares-Discriminant Analysis (PLS-DA) can be conducted. The PLS associates a data matrix containing independent variables and samples to a matrix containing dependent variables (or measurements of response) and e.g. can be used for time series studies. The Discriminant analysis helps finding the position of a discriminant surface that best differentiates classes. In neural networks a training set of data is used to develop algorithms capable of coping with the structure and functional complexity of the data "learning", operating at three or more layers i.e. input level of neurons (variables), one or more hidden layers of neurons, adjusting the weighting functions for each variable, and an output layer that designates the class of the object or sample. A probabilistic technique such as CLOUDS (Ebbels, Keun et al. 2003) (will be discussed later) has important advantages over methods such as k-nearest neighbour, linear discriminant analysis, because unlike these classifiers, the classification is not absolute and therefore, due to extra information (i.e. the class probabilities) there is more room for e.g. examining hypothesis and interpretation.

### 1.3.1. CLOUDS-Overlap

The standard or basic the CLOUDS model (Ebbels, Keun et al. 2003) is a supervised probabilistic approach and non-neural implementation of a classification technique, based on the Parzen density estimator (Parzen 1962). This which implemented into metabonomics initially using probabilistic neural networks (Holmes, Nicholson et al. 2001; Dyrby, Baunsgaard et al. 2005; Smilde, Jansen et al. 2005; Timothy M. D. Ebbels 2006). When modelling complex multi-dimensional distributions, the CLOUDS gives a probabilistic rather than absolute class description of the data by summing Gaussian densities at each training set data point. The potentials such as choosing different smoothing parameters ($\sigma$) makes it is particularly amenable to inclusion of prior knowledge such as uncertainties in the data descriptors. Also, unlike projection methods, it is relatively insensitive to outliers, yet readily detects them as samples with consistently low membership probabilities across all classes (Ebbels, Keun et al. 2003).

Since the classifications are not absolute, CLOUDS brings key improvements such as the uniqueness and confidence diagnostics.

The probability of any data object with feature vector x belonging to class A is (Equation 1):

**Equation 1**

$$p_A(\mathbf{x}|\mathbf{x}_{i \in A}) = \frac{1}{N_A(2\pi\sigma^2)^{M/2}} \sum_{i \in A} \exp\left(\frac{-|\mathbf{x} - \mathbf{x}_i|^2}{2\sigma^2}\right)$$

where σ is a parameter defining the smoothness of the probability estimate and $x_i \in A$ denotes the training data for the NA objects in class A and must be set based on known (priori) information.

The extended version of CLOUDS model is able to provide a measure of similarity between groups of data points using Equation 2

**Equation 2**

$$O_{AB} = \frac{1}{N_A N_B (4\pi\sigma^2)^{M/2}} \sum_{i \in A} \sum_{j \in B} \exp\left(\frac{-|\mathbf{x}_i - \mathbf{x}_j|^2}{4\pi\sigma^2}\right).$$

In a nutshell based on the formulae, each data point from class A measures how far is from all data points of class B. Distances based on σ exponentially decreases probability of being in the same class. Sum of the probability give a estimate of overlapped probability of the two classes A and B. The similarity is between classes A and B is defined by normalising overlap integral (Equation 3), thus similarity measure ranges between 0 and 1, describing the situation when there is no overlap to the situation when two clouds are identical, respectively (Timothy M. D. Ebbels 2006).

**Equation 3**

$$S_{AB} = \frac{O_{AB}}{\sqrt{O_{AA} O_{BB}}}$$

### 1.3.2. SMART scaling

SMART analysis is a modelling strategy based on homothetic geometry (By translation and scaling without rotation, SMART analysis transforms data linearly and thus coinciding trajectories share homothetic geometry.), Together with PCA, SMART was applied to a number of toxicological studies including inter-laboratory variations in (Keun, Ebbels et al. 2004) hydrazine response for visualizing similarity of multivariate responses. These in turn facilitate the interpretation of metabonomic data such as comparative study of endogenous metabolic response to toxins and could be implemented as a mechanism to classify unknown treatments based on responses known treatment (Keun, Ebbels et al. 2004).

### *1.3.3. Commonly used terms in statistical pattern recognition*
#### 1.3.3.1. Cross-validation
Cross-validation is a method to estimate generalization error based on re-sampling i.e. the process of holding aside some training data which is not used to build a predictive model and to later use that data to estimate the accuracy of the model simulating the real world deployment of the model. In k-fold cross-validation,
data is divided into $k$ subsets of approximately equal size and then the model is trained $k$ times, each time leaving out one of the subsets, which is used to compute the error.
#### 1.3.3.2. Classification vs. regression
Classification and regression are standard statistical tools for reconstructing a source (or its attributes) from noise-corrupted data. Classification predicts categorical class labels (discrete or nominal), classifies data (constructs a model) based on *the training set* (The set of instances used for model construction) and the values (class labels) in a classifying attribute. The model can be presented as e.g. classification rules, decision trees, or mathematical formulae. Next, if the accuracy of the model is acceptable, the model can be used to classify future or unknown objects. The accuracy of the model can be estimated by comparing the known label of a *test sample* with the classified results from the model. The percentage of test set samples correctly classified by the model is called the accuracy rate. The test set must be independent of training set;

otherwise the result is biased i.e. over-fitting occurs. Regression is a data analysis technique which is used to build predictive models for continuous prediction fields. It determines a mathematical equation that minimizes some measure of the error between the prediction from the regression model and the actual data.

#### 1.3.3.3. Supervised vs. unsupervised methods

In unsupervised learning, or clustering, the goal of the analyses is to uncover trends, correlations, or patterns, and no assumptions are made about the structure of the data. In other words, a model is built without a well defined goal or prediction field. In supervised learning or class prediction, knowledge of a particular domain is used to help make distinctions of interest and e.g. involves learning a function y = f(x) from training examples of the form (x, f(x)).

### 1.3.4. Pattern recognition techniques

Like any other omics approaches such as genomics and proteomics, metabonomics also inevitably produces a massive information requiring extraction of an abstract, intelligible and most meaningful representation out of data to generate useful new knowledge with mechanistic and descriptive power(Goodacre, Vaidyanathan et al. 2004). This will help also organise the experimental driven observations into a rational scheme. Pattern recognition techniques also might be used to determine the presence or the extent of any meaningful and significant difference between test and control models. Principle component analysis (PCA) and PCA based models are the most commonly used pattern recognition techniques in metabonomic investigations.

### 1.3.5. Principle component analysis (PCA)

Pearson initially formulated PCA (Pearson 1901) and then Fisher and MacKenzie (R. Fisher 1923) outlined NIPALS algorithm, which later applied to chemometrics (Wold, Esbensen et al. 1987; Wichern 1992). PCA properly handles matrices with more variables than observations and potentially containing noisy and highly collinear data. This technique can be used for summarizing and data reduction, visualizing, preprossessing, classification and discriminant analysis, variable selection, predication, unmixing variables, and finding quantitative relationships among the variables (Wold,

Esbensen et al. 1987; Eriksson, Antti et al. 2004). PCA estimates the correlation structure of the variables and in general any data matrix can be simplified by PCA, and e.g. it can be used to build a descriptive model of how a biochemical system behaves. PCA extracts the dominant object pattern and providing more comprehensible information as a new set of variable called scores, which are weighted sum of the original variables. The link between scores and original variables are the 'loadings' which define the weight or influence of the original variables on the pattern described by scores.

The most common use of PCA is probably its ability to reduce the dimensionality (in multidimensional data matrices) of data into a simplified graphical plot(s) in a space defined by components.

### 1.3.5.1. Some Important features of PCA:

- Each component is a linear combination of original variables
- Components are orthogonal and therefore each component describes the maximum variation not included in the previously calculated components
- PCA plot captures and displays the most dominant patterns in the data matrix it is generally assumed that directions of maximum variance is the representation of maximum information

## 1.4. Metabolic trajectories

Although xenobiotic metabolism is the product of probability not design and therefore metabolic pathways should not be considered a sort of industrial production line, however, pathway undoubtedly can be considered as a valuable and useful diagrammatic summary (Wilson and Nicholson 2003). Analogously, an organism's response to a stimulus, can be summarised using the metabolic trajectory (i.e. the path that it follows over time through an abstract metabolic space, obviously not signalling pathway). "The term trajectory means a path, especially when parameterized by time, and the aims of metabonomic investigation could be interpreted as the attempt to define the metabolic trajectory resulting from a particular intervention: the coordinate system defines the set of metabolites that is measured, and the position on the path indicates the current abundance or flux of the species of interest" (Keun, Ebbels et al.

2004). Varying metabolites along time, whether enriched or depleted, as well as dynamic covariations among them, which are of particular interest will determine the direction and distance travelled when the trajectory is visualised. Monitoring perturbations of metabolites (e.g. in serum or urine), induced in response to the natural cycle of a toxic process i.e. peak (maximum potential damage) and recovery (potential regeneration) reflects a pattern characteristic of a specific pathological process (e.g. mechanism of toxicity). This is influenced by various factors such as the nature of the toxic agents, dosage, body's response and the extent and the number of organs involved as well as variations between individuals.

Unavoidable variations arise during experiments (e.g. variations due to individual responses, unavoidable inter-laboratory differences, etc). Therefore, representation of an organism's metabolic response to treatment using metabolic trajectories irrespective of baseline values and overall magnitude facilitates the incorporation and comparison of metabonomic data sets. This can improve the accuracy and precision of classification models (Keun, Ebbels et al. 2004).

### 1.5. *Simulation and modelling:*

During experimental design there are different means and specific considerations and measures, which can be are taken to minimise and detect any systematic bias experimental errors during interpretation. It is not possible to create an ideal experiment, which can eliminate any intervening factor other than those of interest. Therefore, in many circumstances, especially for testing methodologies, it is very useful and logical step to simulate data and produce computationally simulated data to be able to take control of all variables and observe the exact effect of any particular factor on the performance of the model and the final results. Computer modelling can be used for hypothesis generation and prediction, system level insights.

### 1.6. *Aims:*

As mentioned in earlier sections, the exact trajectories followed within a given treatment group differ due to variations between individuals. Although different techniques such as SMART may at least partially balance the variations, more focused alignment techniques and protocols as well as novel bioinformatics tools which take into account the natural variations in speeds and magnitudes of responses between individuals are required so that characteristic or most probable trajectory of a treatment can be defined, which in turn can be used in interpretation and comparison toxicological responses in a wider range of metabonomic studies.

By adoption of the two techniques (i.e. CLOUDS-Overlap and SMART) and a new approach based on the analysis of sum squared errors and testing the results on both experimental and computationally simulated data, this study aims at outlaying most accurate methodology and protocol for trajectory alignment with a view to defining the most probable trajectory of species- and treatment-specific biological response.

*Aims and objectives can be summarised as:*

1. Clearly defining trajectory alignment concept
2. Developing reliable computational tools as well as mathematical methodologies for simulation of metabolic trajectories and modelling which can be used for testing trajectory alignment
3. Developing alignment strategy based on the CLOUDS Overlap method
4. Developing alignment strategy based on the standard error analysis
5. Exploring potential capacities of standard methods such as PCA which are commonly applied in metabonomics
6. Comparing pros and cons each technique especially from trajectory alignment perspective and possibly providing recommendations for applicability and suitability of each technique under different conditions and circumstances

# 2. Materials and Methods:

Alignment in general can be defined as the process of adjusting parts so that they are in proper relative position. This can help examine if two or more shapes are similar if they are aligned properly. Similarly, 'toxin likeness' of one compound to another can be described by comparing the metabolic trajectories (connected the data across all time points) resulting from organisms' response. In this study the trajectory alignment is referred as the process of extracting maximum similarity via calibrating the scale of trajectories.

According to homothetic geometry hypothesis (Keun, Ebbels et al. 2004), the shape of a 'metabolic trajectory' is indicative of the underlying biological cycle. Deformation of the shape or rotation of the trajectory is not allowed so that the integrity of the biological concept is maintained. Thus, alignment analysis of trajectories can rely merely on similarity measures of metabolic trajectories along a range of scales as well as linear translation of data points. Consequently, the best trajectory alignment will be expressed by relative scales of two (or more) trajectories when the maximum possible similarity occurs.

Thus, in principle, trajectory alignment is indeed combination of translation and scaling optimisation scheme aimed at exploring maximum potential similarity among trajectories based on the assumption that maximum similarity is not a coincidental phenomenon, but indicative of a unique mechanism with diverged manifestations in form of metabolic trajectories owing to a number of factors such as biological time differences between organisms. This conjecture also has been examined by cross validating (correlating) the similarity measures of metabolic trajectories obtained from NMR spectroscopy of urine samples of the mice with the severity of histo-pathological damages in liver samples taken from the same mice euthanized after the urine samples had been taken (Hector C. Keun 2006).

In this section, the overall structure of the methodology is outlined and more detailed explanations are provided under the relevant subheadings.

Conceptually, alignment of metabolic trajectories in this study is comprised of two major components:

1. Similarity measures
2. Acceptable transformations

Similarity measures between trajectories were detailed by employing two following techniques:

a. The extended version of the CLOUDS (Ebbels, Keun et al. 2003), so called the Overlap technique (Timothy M. D. Ebbels 2006)
b. A new method based on Sum of squared Error Analysis (SEA)

Methodologically, similarity analysis was formed based on two major approaches:

1. Local approach: In this approach, in calculation of similarities, not all of data points belonging to each trajectory are necessarily taken into account.
2. Global approach: This approach, which employs the CLOUDS and the Global SEA technique, takes account of all data points belonging to all trajectories during similarity calculation between two (single or group of) trajectories

Figure 3 Schematically presents the entire trajectory alignment process in this study. SMART can be considered as scaling scheme.

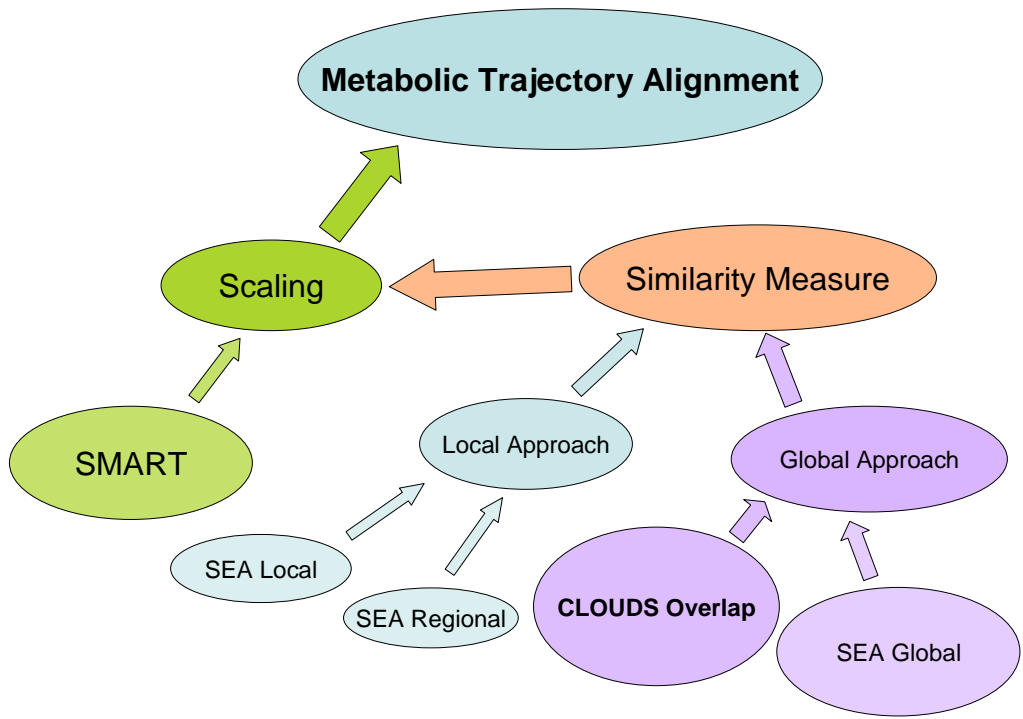

**Figure 3 Overview of metabolic trajectories alignment**

For alignment purposes, a SMART- (Keun, Ebbels et al. 2004) based method has been employed. Standard SMART technique is 'rigid' in that sense that it scales trajectories into their maximum dimension only. Since metabolic response irrespective of baseline values and overall magnitude, defines the mode of response of the organism to treatment (Keun, Ebbels et al. 2004), SMART pre-treatments or linear scaling does not affect the results according to homothetic geometry. In this study SMART is applied on each (single of group of) trajectory initially to 'normalise' them so that trajectories become more comparable to each other. Various scales are applied to 'normalised' trajectories and each time similarity between trajectories is measured. Therefore, this is a SMART based approach which allows various scales to be used and therefore can be considered 'flexible' compared to standard SMART. For this reason and for simplicity this method will be referred as F-SMART (i.e. Flexible SMART). Two differences between F-SMART and SMART are:

- Additional scaling step following standard SMART
- Translation of trajectories into their median instead of one time point (usually first or second) which is done in standard SMART

Modelled trajectories were created using computationally generated data. During entire analysis, (MathWorks Inc.Version 7.0.0.19920 R14) was used for programming, data handling, statistical analysis, developing and testing computational models as well as data simulations.

## 2.1. PCA

PCA was carried out to reduce dimensionality of both the NMR and a simulated data, using a MatLab subroutine, which was developed based on the NIPALS algorithm (R. Fisher 1923; Erriksson 2001) To examine the subroutine's performance, a 100 × 240 matrix (100 dimensions and 240 variables) was generated, using random normal variates. Subsequently, coordinates along $4^{th}$, $10^{th}$, and $36^{th}$ dimensions were multiplied by 5, 10, and 20 respectively. Dimensions and scale factors were chosen arbitrarily. Scores, residuals, and loadings of the data matrix computed using the developed PCA subroutine (first 4 principle components). The original coordinates as well as the residuals of the same dimensions (i.e. $4^{th}$, $10^{th}$, and $36^{th}$) were visualised using black, blue, green, and red colours respectively to create 3D image See Figure 4.

Loadings of PC 1 to 4 also were visualised using the same colour scheme corresponding to the residuals of components (i.e. blue, green, red) for loadings of PC 1 to 3 respectively and brown for loadings of PC 4. According to the loadings shown in Figure 4, as expected, the first component (PC1) is mainly influenced by $36^{th}$ dimension (the biggest variation i.e. multiplied by 20), PC 2 by $10^{th}$ (multiplied by 10) and PC 3 by $4^{th}$ dimension. PC 4 Loadings is not pronounced at towards any particular dimension. These features explain that PCA code is working.

Since loadings are normalised, therefore existence of one extreme value automatically compresses the other peaks. As shown in the Figure 4, loadings of first 3 PCs describing 3 outliers, contain one prominent peak on almost a straight line (other peaks are shortened), whereas loadings of $4^{th}$ PC show very noisy pattern and there is not a prominent peak. Because loadings reflect the influence of the original variables on scores, therefore it can be intriguing idea to use loadings plot to determine the number of components, which sufficiently explain a data set. One example of the standard methods is 'scree graph', which can be used to determine rank of the PCA model of a matrix (Jolliffe, Uddin et al. 2002). Figure 4 extracted from a published paper (Jansen, Hoefsloot et al. 2004) is displayed as an example for use of 'scree graph' to determine the number of principle component in which principle components and Eigenvalues were plotted on x and y axes respectively. For instance authors in this study have chosen 3 principle component explaining explains 66% of the variation in the data (Jansen, Hoefsloot et al. 2004)

At the bottom there are 3 different views of visualisation of a 100 by 240 data matrix with maximum variations along 3 dimensions

Original variables shown with black colour and the residuals of PC1, 2, and 3 with blue, green, and red colours respectively.

On the right, loadings of PC1 to 4 are shown with blue, green, red, and brown colours.

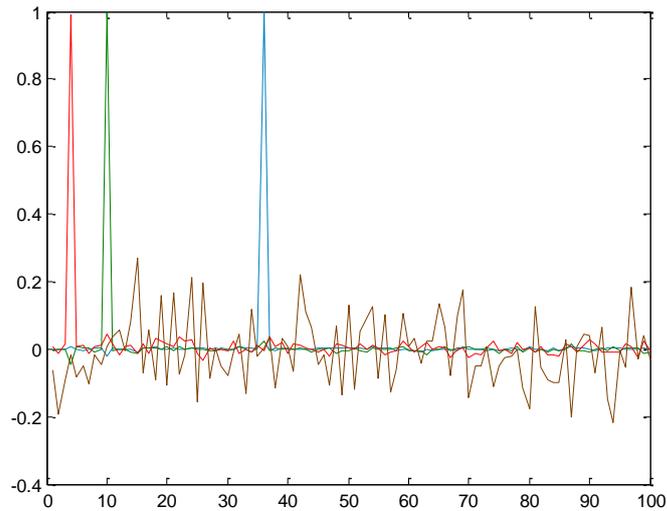

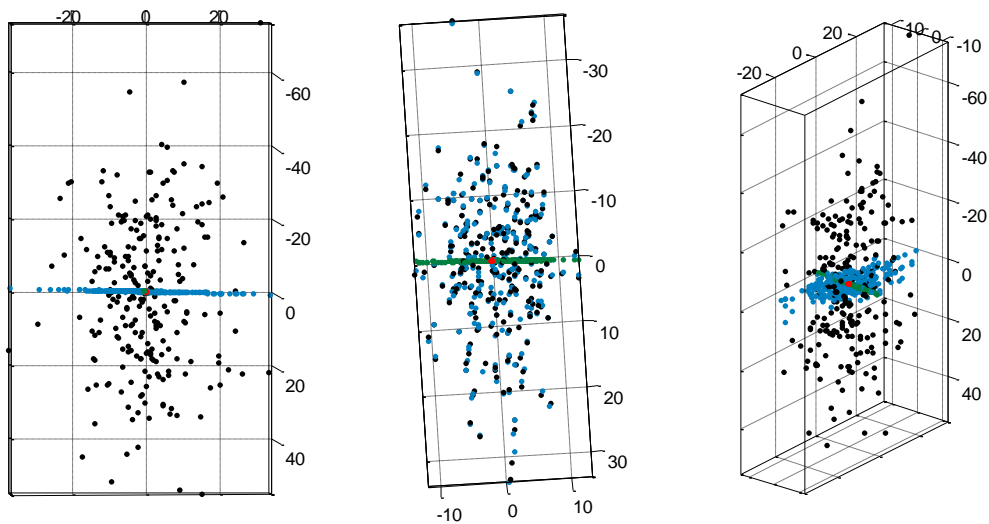

**Figure 4: PCA, first 3 components, loading plots**

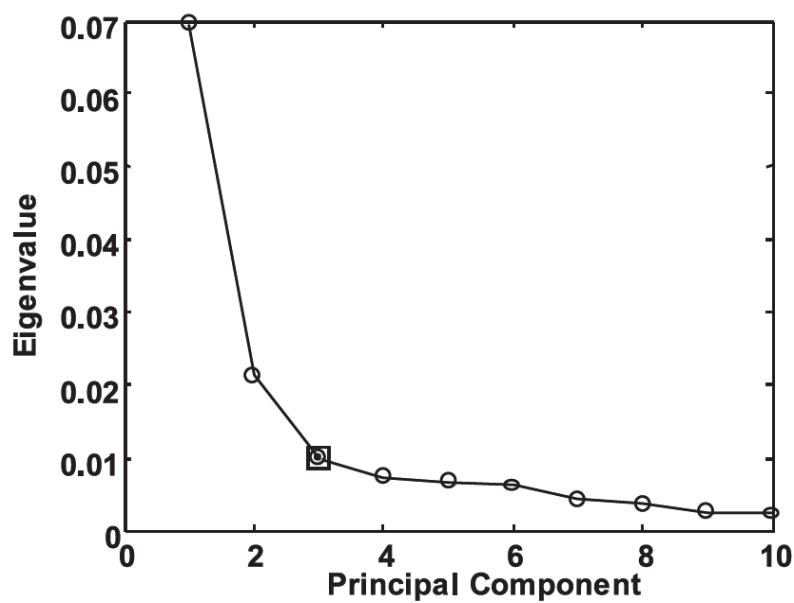

**Figure 5: Scree graph (extracted from published paper(Jansen, Hoefsloot et al. 2004))**

## 2.2. *Data simulation:*

Metabolic trajectory is categorized as time-series analysis i.e. there is an ordered sequence of values of variables derived from samples taken at equally spaced time intervals. To model a metabolic trajectory, identification of the underlying forces and structure that produce the observed data is essential.

### 2.2.1. *Metabolite concentration function*

The concentration of a metabolite (serum or urine) which is affected during biological response to a xenobiotic agent follows a path, which is typically an asymmetric (a positive skew or skewed to the right) curve beginning with a sharp rise up to a peak (e.g. damage due to toxicant) and then relatively smooth decline to the original level (recovery or regeneration). This curve with a long tail in the positive direction can be graphically visualised using coordinate systems in a way that for instance time and the metabolite concentration are represented by X and Y axes. Figure 2 A illustrates a modelled curve analogous to biological response.

The following formula (coined as Metabolite Concentration Function, MCF) can be used to simulate the positive skew curve, an analogous to the natural cycle of the biological response:

**Equation 4**

$$f_{(t)} = m\left(\frac{\beta \times t}{\alpha}\right)^{\alpha} e^{\alpha - \beta t}$$

, where

$f_{(t)}$     is simulated level of the metabolite concentration in urine or serum,

$m$     is the peak of the concentration (biological damage),

$\alpha$ and $\beta$     determine the skewness i.e. degree of asymmetry,

$t$     represents time, and

$e$     is the exponential.

According to the formula, it possible to determine the overall magnitude (the biggest possible number which $f(t)$ can reach, determined by $m$) and the skewness (asymmetry in intensity of rise and fall of the curve analogous to a natural cycle of biological response e.g. speed of damage and recovery in real world situation) of curve through manipulating $\alpha$, $\beta$, and $m$.

Similar to real world situation, where metabolites' concentrations can only be measured at specific time points (i.e. when the samples are taken), $f_{(t)}$ is only measured at distinct distances of $t$ from the origin, where the distance is analogous to time in real world situation and $f(t)$ represents the metabolite concentration measured at that time point $t$. Thus, differences in the speed of response (assuming identical pattern of response in terms of. skewness and overall magnitude) between animals A and B when B's response is faster than A's response, can be simulated by considering longer distances of $t$ from the origin for animal B compared to animal A while treating $f(t)$ in such a way as if they were taken exactly at similar distances of $t$ from origin or $t$ during analysis. In other words, a metabolite concentration in animal B is travelling longer distances of the same path (the curve) at a given time. Thus, using the above formula, the concentration of a given metabolite in animal A at $t_i$ is calculated as $f(t)$, whereas the concentration of a given metabolite at the same time point in animal B is calculated as $f(t)$, where $t_i = t_j + \Delta_i$.

By manipulating values assigned to instances of $\Delta_i$ (The gap between the distances of $t_i$ and $t_j$ from origin, representing speed difference in real world), different speeds can be simulated. For instance if values assigned to $\Delta$ are in increasing order, it means the animal B's response is faster with an accelerating speed. Figure 2 A and B illustrates simulation of speed difference (or biological time difference).

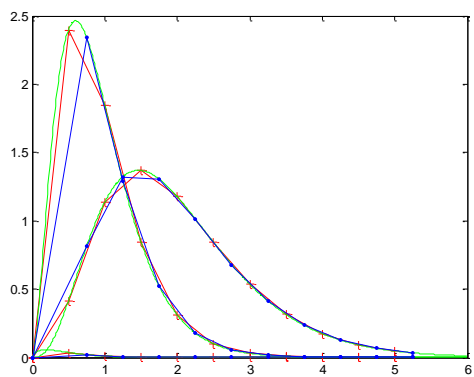

A: Green line: simulation of three different metabolite perturbations in three different model biological responses (asymmetric positively skewed curves) along time. Peaks and tails are analogous to cycle of biological damage and regeneration or recovery. Three curves or metabolites are considered as three different dimensions (chemical shifts) in metabonomics analysis. Blue and red lines, which respectively represent two different animals, simulate measure of concentrations of the related metabolites (samples taken at different stages) at two different speeds (the red colour represent the biological response with delay)

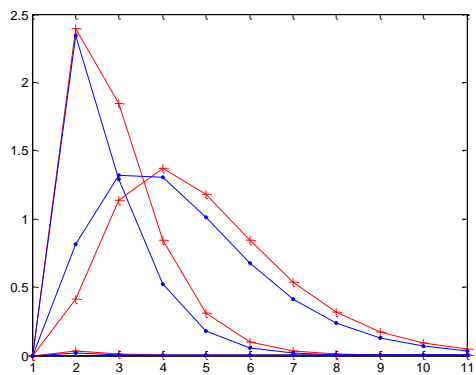

B: The speed difference between the two modelled animals for simulated metabolite concentrations (blue and red for normal speed and delayed respectively) are better displayed after the coordinates are superimposed

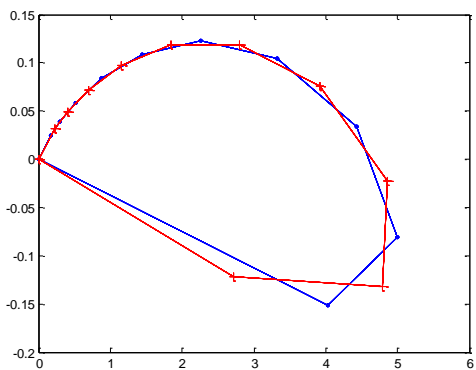

C: Visualisation of the metabolic trajectory of biological responses related to two modelled animals using two components (PC1 and PC2). After PCA using in-house routine based on NIPALS algorythm, faster onset of the modelled animals with blue colour is displayed in the trajectory

**Figure 6: Simulation of biological time difference**

During the analysis of metabonomic data, each measured metabolite (or chemical shift) is considered as one variable. Typically, there are always a number of sets of covarying variables, which are detected by pattern recognition techniques such as PCA. For this reason, during modelling of each animal, 10 different positive skew curves were created. To generate each curve, each time random normal variates were assigned to $\alpha$, $\beta$, $m$. In modelling, three variables coined A_Range, B_Range, and M_Range were considered to define the ranges of random normal variates $\alpha$, $\beta$, $m$, respectively.

For $t$ a range of 10 values in an increasing order with equal interval were used to simulate 10 samples taken from every animal at 10 different time points, except the times when generating different speeds were intended and then intervals were not equal. Subsequently, each modelled metabolite signal was replicated 10 times with slight variations within a defined range of variance so that ultimately 10 sets of dimensions, each set containing 10 covarying dimensions with equal variances were created. Since during modelling each curve (simulating biological response), $f_{(t)}$ was calculated at 10 different $t$ values, therefore, every modelled animal was defined by a $10 \times 100$ matrix (10 rows for time points and 100 columns for dimensions or modelled metabolites). As mentioned earlier, 100 dimensions structurally could be divided into 10 sets, each set containing 10 strongly covarying metabolites.

The following formula explains how simulated concentration levels of a metabolite along time using MCF are replicated to generate a set of relatively similar values within a defined variance:

**Equation 5**
$$u_{i(t)} = (1 + P_i) f_{(t)}$$

, where $f_{(t)}$ is metabolite concentration function, i.e. Equation 1, for a given time point ($t$). In Equation 5, $i$ represents each replicate (in our model there are 10 replicates i.e. $i : 1 \rightarrow 10$) termed as 'Realisations', and $P$ is a random normal variate and therefore $u_{i(t)}$ is a new coordinate and $P_i f_{(t)}$ is the amount of deviation from $f_{(t)}$ a variable in our modelling termed 'Deviation'. These new coordinates are in fact simulating of biological noise which are normally distributed. Therefore, the range of biological noises at metabolite level could be set to a desired amount by defining the range of $P_i$. Figure 3 A illustrates 2 curves (Green lines), 10 samples from the two curves (Red lines) and 5 realisations (Blue lines) created using above formula i.e. Equation 2, where $i = 5$ and $P_i$ is simplified from a uniform distribution between -0.05 and 0.05. Figure 3 B illustrates the trajectories of the curves with their corresponding colours.

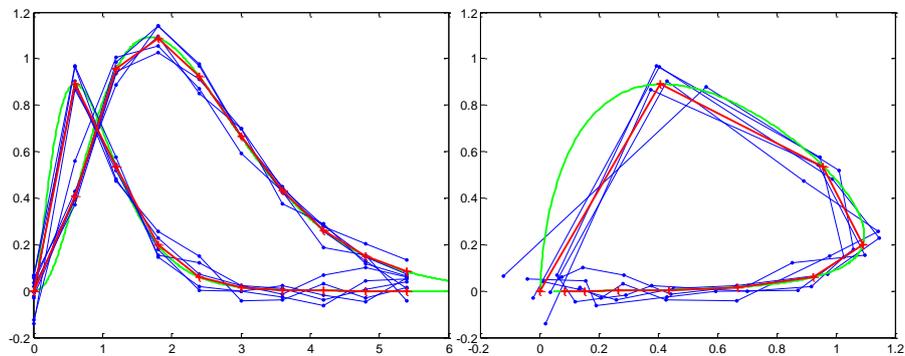

A (left): Green lines: simulate metabolic biological response; Red lines: illustrate the levels of simulated metabolite concentrations at specific (time) points; Blue lines: illustrates 5 realisations out of each simulated concentration (σ= 0.05)
B (right) Trajectories of the curves with corresponding colours

**Figure 7: Simulation of biological noise**

With regard to the simulation of noise or 'Deviation' variable, depending on the biological question, there is a useful practical point to be aware of. During the analysis, if modelled trajectories are going to be scaled or normalised and compared, probably it is a good practise to set the level of noise 'Deviation' proportionate to the magnitude of the original modelled curves. This matter will be mentioned later when application of SMART technique(Keun, Ebbels et al. 2004) is being discussed.

To generate a similar animal model but with small normal variation, the following equation was used which is similar to equation 2:

**Equation 6**
$$g_{i(t)} = (1+Q) \times u_{i(t)}$$

, where $g_{i(t)}$ are new coordinates with $Qu_{i(t)}$ variation from $u_{i(t)}$ and $Q$ is the random normal variate, whose range has a multiplicative effect on variations of trajectories of a modelled animal and in our study has is termed the 'Deviation between animals' variable. A similar approach was used to define 'Deviation Between Animal Groups' variable, using random normal variate with multiplicative effect on all variation of a group of modelled animals.

## 2.3. Trajectory Similarity Metrics:

### 2.3.1. CLOUDS – Overlap:

The CLOUDS Overlap (Timothy M. D. Ebbels 2006) (detailed in the introduction) was used as a tool to measure similarity between trajectories.

#### 2.3.1.1. Smoothing factor (σ) optimisation

Initially maximum likelihood, using cross-validation (See introduction) implemented in the accompanied programme for the Overlap was tested. Figure 8 shows the plot in the optimised σ is 1.2250 based on all the experimental data used in this study. Then, because the exact value of σ is not critical, after preliminary investigations on both experimental and simulation data, σ=1 was found generally suitable and used as default throughout the study, unless explicitly stated otherwise.

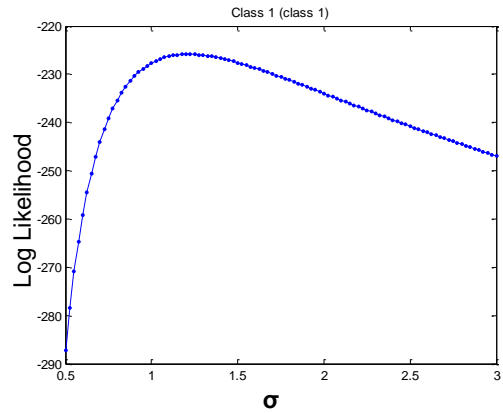

**Figure 8: Sigma optimisation using experimental data**

### 2.3.1.2. Secondary peaks in similarity plot

Probably one of the pitfalls of the trajectory alignment via local approach in general and overlap technique in particular is when data points approach each other as the whole shapes expands due to scaling. The best example of this situation can be shown in grid data points where data points are lined up in similar directions yielding high similarity measures during scaling and consequently e.g. a number of secondary peaks appear if the similarity is plotted against scale factor using CLOUDS-Overlap. Figure 9 illustrates the same effect and also showing how different σ values can affect similarity measures.

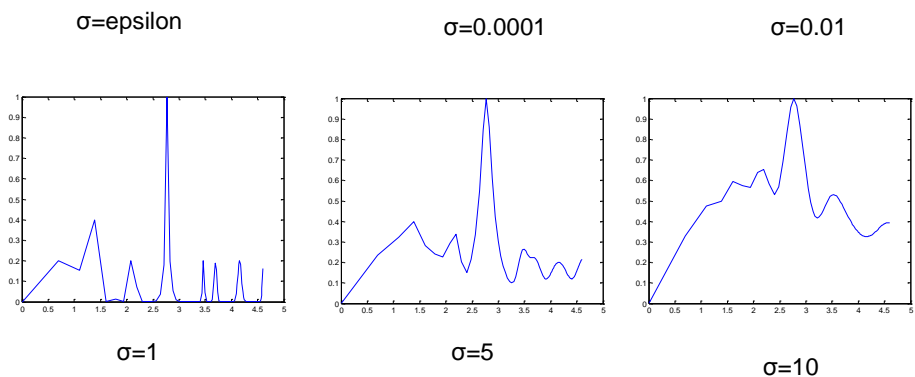

**Figure 9: Different sigma values and similarity plot on grid data**

### 2.3.2. Squared Errors Analysis (SEA):

Probably this is the first time that squared errors analysis is introduced to metabonomics. Squared errors analysis is introduced as implementation standard sum squared sum errors SSE which coined as SEA-local, Goodness of fit for two trajectories (local and regional), and SEA-Global and each of the might address biological questions from different perspectives:

## 2.4. Trajectory Similarity Metrics via Local Approach:

Principally, this approach has been considered assuming it will be able to cope with some specific problems rising during interpretation of metabolic trajectories. 'Biological time' difference is a major factor which can affect interpretation of metabolic state if it is not properly handled by the employed methods. There are also other common practical issues making interpretation of the trajectories more difficult such as missing data and inter-laboratory mismatch in number of time points when biological samples are taken.

What characteristically differentiates local approach in trajectory similarity metrics from global is that in local approach not all data points of trajectories, which are subject to similarity analysis are necessarily taken into account for actual similarity calculations. In other words, out of each trajectory, only a group of selected data points geometrically confined to some mathematical boundaries are considered for similarity measurement. The ways these boundaries are drawn and also the ways similarity calculations are made vary depending on the chosen technique. Techniques, which can be used to measure similarity via local approach, are:

1. SEA-Local
2. Goodness of fit for two individual trajectories - Local
3. Goodness of fit for two individual trajectories – Regional

In principle it is possible to use these elements (defining boundaries and mathematical calculation) interchangeably (i.e. defining mathematical boundaries based on one method and calculating similarity using the other method) which can probably provide further alternatives for similarity-metrics.

### 2.4.1. SEA Local:

Like any other time series analyses, study of metabolic trajectory accounts for the assumption that data points, taken over equally spaced time intervals may have an internal structure (such as autocorrelation, trend). Thus sum of the squared errors between two groups of trajectories, A and B, can address the question whether or not and to what extents they are equal. In this approach one group of trajectories (e.g. A) is used as reference trajectory group, from which expected values can be elucidated, whereas the other one (e.g. B) is used as test or observed trajectory group.

Data points in reference trajectory group are clustered (using Kmeans algorythm) and subsequently the resultant centroids are used to classify the data points of test trajectory group. Any given data point in test group is classified to class $i$ if it has the shortest distance to centroid $i$ compared to the other centroids. During clustering, to increase the efficiency of the programme, medians of the time points of the reference trajectories were used as seeds for clustering.

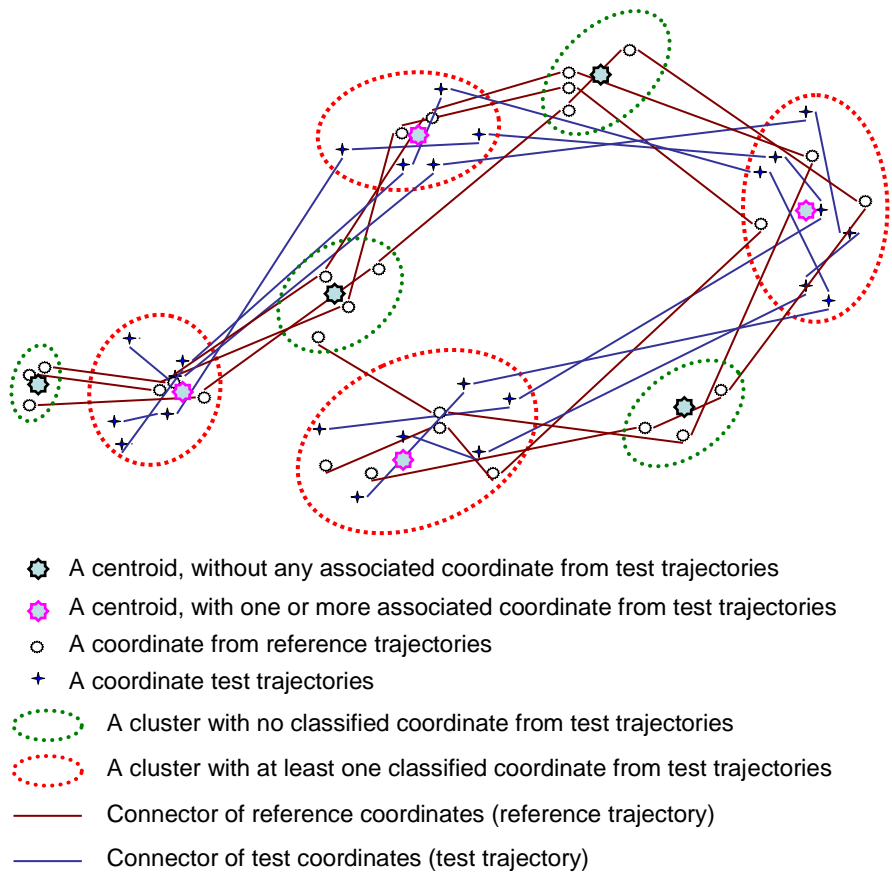

- ⊛ A centroid, without any associated coordinate from test trajectories
- ⊛ A centroid, with one or more associated coordinate from test trajectories
- ○ A coordinate from reference trajectories
- + A coordinate test trajectories
- ⬭ A cluster with no classified coordinate from test trajectories
- ⬭ A cluster with at least one classified coordinate from test trajectories
- —— Connector of reference coordinates (reference trajectory)
- —— Connector of test coordinates (test trajectory)

**Figure 10: Schematic illustration of trajectories from reference and test trajectories and clusters and classes**

Figure 10 schematically illustrates 3 trajectories of hypothetical reference (observed) and 3 of test data coordinates. Coordinates of reference trajectories develop 8 clusters and naturally there are 8 centroids, of which only 4 cetroids are accounted for calculation because each of these four cetroids (shown in red 8-points star) form a class containing at least 1 coordinate from test trajectories (circled by red dashed lines). Each coordinate form test trajectory is classified into the nearest centroid from reference trajectory clusters.

To calculate SEA local, first within scope of each class, sum of the squared distances of test trajectory coordinates from the associated centroid is calculated and divided by median of the squared distances of reference trajectory coordinates from the centroid. Calculating this way, sum of the values derived from all classes is considered as SEA local value. See Equation 7 below and the following algorithm:

**Equation 7: SEA Local**

$$\chi^2 = \sum_{i=1}^{k} \frac{\sum_{j=1}^{m_i} |\underline{X}_j - \underline{C}_i|^2}{median\left(_{l=1}^{n_i} |X_l - C_i|\right)^2}$$

Where:

$k$ : The number of clusters developed from Ref data set (During clustering, medians of the Ref dataset at each time point were used as seeds of SOM-Kmeans)

$\underline{C}_i$ : $i^{th}$ centroid of Ref data set

$m_i$ : The number of data points from Test data set, classified as class $i$

$n_i$ : The number of data points from Ref data set in cluster $i$

$|\underline{X}_j - \underline{C}_i|$ : Distance of $j^{th}$ data point from $i^{th}$ centroid within the scope of $i^{th}$ class

$|\underline{X}_l - \underline{C}_i|$ : Distance of $l^{th}$ data point from $i^{th}$ centroid within the scope of $i^{th}$ cluster

*Algorithm:*

*Step 1:* Cluster Ref data set, while the medians of coordinates of Ref data at each the time point are initialised as seeds for clustering

*Step 2:* Classify each data point from Test data set to the nearest centroid

*Step 3:* Within scope of cluster $i$, calculate median of the distances of Ref data points from the centroid

*Step 4:* Within scope of class $i$, calculate the sum of the squared distances of every Test data point from the centroid and divide by median of distances for cluster $i$ from step 3

*Step 5:* Sum the results of calculations in step 4 for all classes

This formula is directional meaning that the results might change if reference and test trajectories are replaced.

### 2.4.2. Goodness of Fit for two Trajectories (GFT) Local:

GFT-Local is a specific form of SEA-Local where there are only two individual trajectories (reference and test). In this situation, because expected values cannot be measured, therefore, Ref data points are treated as centroids (analogous to centroids SEA-Local) and the Test data points are classified to them according to their distances as described in Equation 8 below:

**Equation 8**

$$\Delta^2 = \frac{N}{M^2} \sum_{i=1}^{n} \sum_{j=1}^{m_i} \left| \underline{X}_{j(test)} - \underline{X}_{i(ref)} \right|^2$$

where:

$N$ : Total number of the reference trajectory time points

$M$ : Total number of the test trajectory time points (i.e. $\sum_{i=1}^{n} m_i$)

$n$ : Total number of classes

$\underline{X}_{i(ref)}$ : $i^{th}$ Ref trajectory data point which centres a class (class $i$), i.e. compared to the rest of the data points, it has the shortest distance to at least one of Test trajectory data points

$\underline{X}_{j(test)}$ : $j^{th}$ data point from Test trajectory within the scope of class $i$

$m_i$ : Total number Test trajectory data points within class $i$

This formula is directional meaning that the results might change if reference and test trajectories are replaced.

### 2.4.3. Goodness of Fit for two Trajectories (GFT) Regional:

GFT-Regional is an extended version of GFT-Local which performs in both directions and measures a wider domain of data point (i.e. the boundaries around the classes are likely to expand, the reason why 'regional' name is chosen in contrast to 'local').

For instance if A and B are two trajectories, first data points A are classified to B data points (**Error! Reference source not found.** on the top, where schematically classes circled by red dashed lines). Next, data points B are classified to data points A (**Error! Reference source not found.** in the middle classes circled by blue dashed lines).

As per **Error! Reference source not found.**, GFT-Regional is measured by summing the squared distances between data points A and data points B only within merged classes (**Error! Reference source not found.** in the middle merged classes circled by green dashed lines) excluding redundant distances. **Error! Reference source not found.** illustrates 3 steps

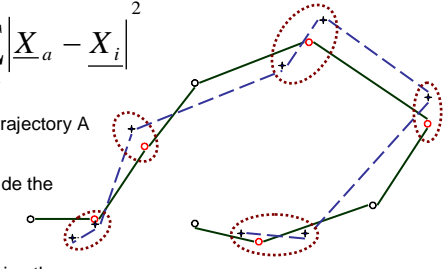

Data points from trajectory A are classified to the nearest data point from trajectory B

$$d_A^2 = \sum_{i=1}^{A}\sum_{a=1}^{m}\left|\underline{X}_a - \underline{X}_i\right|^2$$

+ Data point from trajectory A
○ Data point from trajectory B outside the classes
○ Data point from trajectory B centring the classes
-- Trajectory A
— Trajectory B

Classes A (data points from trajectory A are classified to data points from trajectory B)

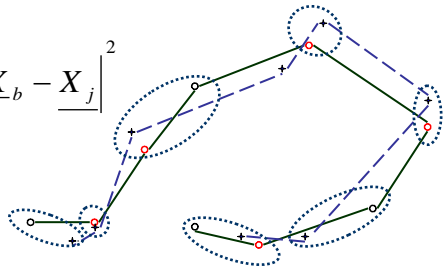

$$d_B^2 = \sum_{j=1}^{B}\sum_{b=1}^{n}\left|\underline{X}_b - \underline{X}_j\right|^2$$

Classes B (data points from trajectory B are classified to data points from trajectory A)

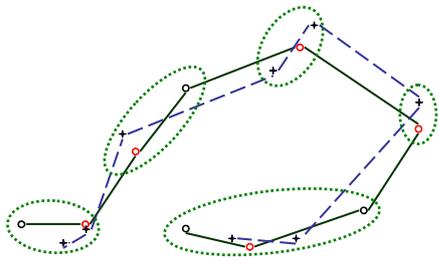

Classes A and B are superimposed

**Equation 9**

$$\delta = \frac{1}{(N_A + N_B)} \left( d_A^2 + d_B^2 - d_{A \cap B}^2 \right)$$

where:

$$d_A^2 = \sum_{i=1}^{A} \sum_{a=1}^{m} \left| \underline{X}_a - \underline{X}_i \right|^2$$

and

$$d_B^2 = \sum_{j=1}^{B} \sum_{b=1}^{n} \left| \underline{X}_b - \underline{X}_j \right|^2$$

and

$A$ : Number of classes when trajectory A is classified to trajectory B

$m$ : Number of data points from trajectory A within $i^{th}$ class

$B$ : Number of classes when trajectory B is classified to trajectory A

$n$ : Number of data points from trajectory B within $j^{th}$ class

$X_a$ : $a^{th}$ data point from trajectory A within $i^{th}$ class

$X_i$ : $i^{th}$ data point from trajectory B centring $i^{th}$ class

$X_b$ : $b^{th}$ data point from trajectory B within $j^{th}$ class

$X_j$ : $j^{th}$ data point from trajectory A centring $j^{th}$ class

$N_A$ : The number of time points in trajectory A

$N_B$ : The number of time points in trajectory B

$d_{A \cap B}^2$ : Sum of squared the redundant distances between data points of A and B after both classifications

$d_A^2$ : Sum of the squared distances of trajectory A data points form their nearest data points in trajectory B

$d_B^2$ : Sum of the squared distances of trajectory B data points form their nearest data points in trajectory A

## *2.5. Trajectory Similarity Metrics via Global Approach:*

As mentioned earlier, this approach takes account of all data points belonging to all trajectories during similarity calculation. Two techniques have been used:

1. CLOUDS-Overlap (Described earlier)
2. SEA Global

### 2.5.1. SEA Global:

In this technique, the squared sum of the all distances between data points of two (individual or groups of) trajectories are calculated. To do this, trajectories should be (at least partially) superimposed by translating them to one of their time points or preferably to their medians or means. A trajectory can be conceived as a polygonal shape and thus it can be assumed that the sum of the squared distances between intersections of shape A and shape B is measured (e.g. in case of two individual trajectories). When calculating the distances in this way, there are some distances which cannot be considered as 'errors', which somehow must be eliminated from the sum of the errors. These 'non-error' distances (NERD) are generated if the line connecting an intersection from shape (trajectory) A to an intersection from shape B crosses over the area which both shapes overlap. This effect can be coined as 'diagonal effect' because the NERDs are similar to diagonals of a shape. Figure 11 schematically illustrates two ideal trajectories in form of two polygons and the connections between the intersections of the two superimposed shapes. The NERDs and 'diagonal effect' are shown with semi-transparent light brown and the 'error distances' with light green colours. The proposed Equation 10 sums the squared error distances after removing the estimated squared distances of NERDs.

**Equation 10: Global SEA**

$$SEA_{Global} = \sum_{i=1}^{N_A} \sum_{j=1}^{N_B} (\underline{X}_i - \underline{Y}_j)^2 - NERDs$$

where:

$N_A$: Number of data points in (an individual or a set of) Test trajectories

$N_B$: Number of data points in (an individual or a set of) Ref trajectories

$X_i$: $i^{th}$ data point of (an individual or a set of) Test trajectories

$Y_j$: $j^{th}$ data point of (an individual or a set of) Ref trajectories

NERDs can be roughly estimated via Equation 11

**Equation 11: Non-Error Distances**

$$NERDs \approx \left(1 + \frac{|N_A - N_B|}{N_A}\right) \sum_{i=1}^{N_A} \sum_{j=1}^{N_A} (\underline{X}_i - \underline{X}_j)^2$$

where:

$X_j$ : $j^{\text{th}}$ data point of (an individual or a set of) Test trajectories

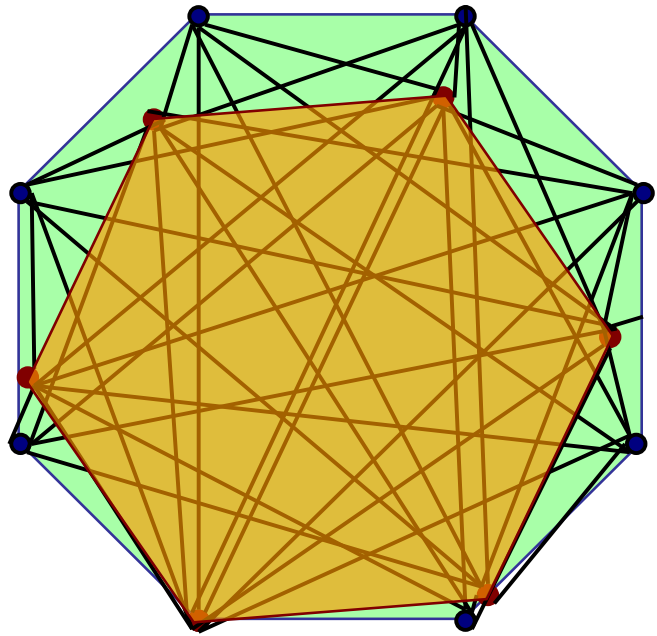

**Figure 11: SEA Global**

## *2.6.* *Trajectory Alignment*

A trajectory and translated and/or expanded or contracted versions of it (homothetic trajectories(Weisstein; Borowski 1989)) share many important features such as the correlations in relative size and direction of metabolite, yet they disregard the initial metabolic and the overall magnitude of a treatment-related effect as well as the exact time of events for interpretation (Keun, Ebbels et al. 2004). Because a rotated trajectory no longer represents important aspects such as direction of the metabolite within metabolic space, so far, rotation has not been generally accepted in metabonomic analysis.

### *2.6.1. Translation*

In metabonomics, trajectories are allowed to be translated because, in many circumstances, the comparison of initial metabolic states and the exact time of events are not of major concern.

Methodologically, trajectories can be translated to:

1. a specific single time point - peripheral translation (PTN)
2. the median/mean - central translation (CTN)

The position of the homothetic centre (HC) of a trajectory moves towards the time point chosen for translation during the PTN, whereas during the CTN, HC moves to the centre. HC (shown in Figure 12) is the focal point where the connectors of corresponding points from homothetic figures meet. Each connector is divided in the same ratio at this point (Weisstein; Johnson 1929).

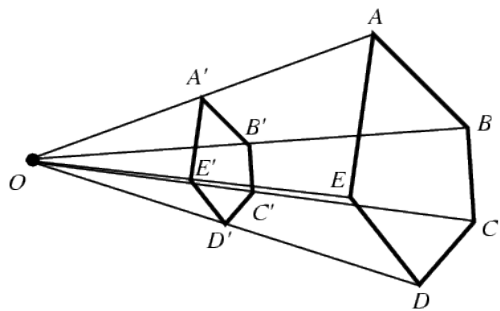

$O$ is the homothetic centre of the homothetic figures $A B C D E$ and $A' B' C' D' E'$

**Figure 12: Homothetic Centre (Extracted from a web resource (Weisstein))**

Whether biological interpretations can directly be influenced by HC position, might require further and specific investigations beyond the scope of this report, however, they are likely to be affected indirectly at least, owing to effect of the HC on similarity measures especially when similarities of two or more trajectories are calculated along a range of scales. Quite often, different trajectories have different directions and if the HC is not centrally positioned, during scaling of trajectories, the data points are likely to be engaged less effectively and at least biased towards the partial overlapping areas. When there is only a partial overlap of trajectories, model scores and residuals are more reliable parameters to compare (Keun, Ebbels et al. 2004). although, loadings can provide a scale-free and baseline independent means (Benigni and Giuliani 1994). Comparison between peripheral and central translations is illustrated in Figure 13. The 'partial overlap' is clearly visible in the figure. Peripheral translation has been applied in metabonomics for alignment purpose in SMART method, which aimed at removing predose differences as major magnitude differences in the data (Keun, Ebbels et al. 2004). It should be taken into account before choosing the translation method depending on the objectives of the study that probably the central translation does not specifically eliminate predose differences

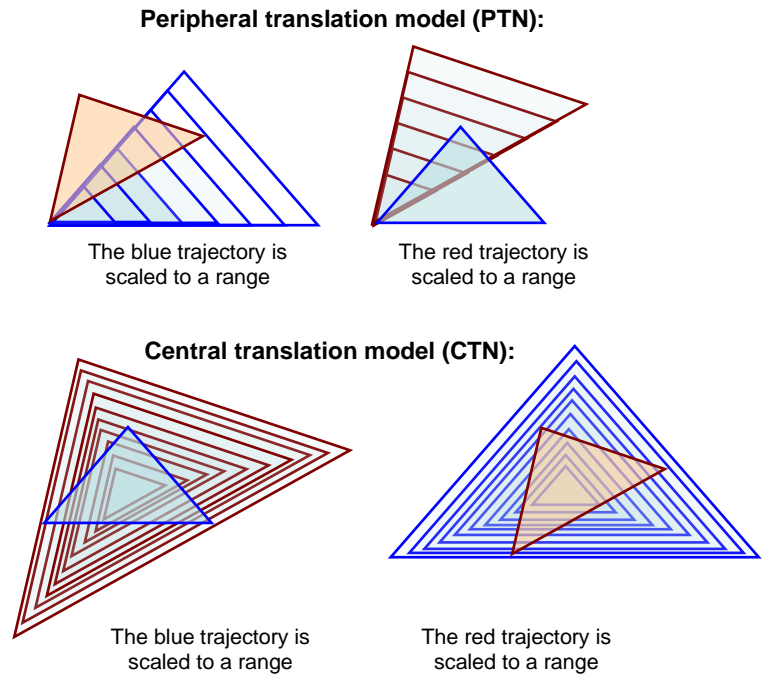

**Figure 13: Peripheral and central translations**

### *2.6.2. Scaling*

In addition to translation, trajectory alignment can be further calibrated through scalar dilation or contraction, disregarding the overall magnitude of response.

#### 2.6.2.1. SMART analysis

SMART analysis can be considered a specific form of alignment, which originally was designed to accurately test for coincident geometry (homothety) of trajectories using PCA. The average spectrum from the 0 h measurement for each study was subtracted from all of the spectra for each study and all time points are divided by the distance of the furthest time point from 0 measurement, assuming that the largest distances provide an estimation of the overall magnitude of the treatment-related response (Keun, Ebbels et al. 2004).

#### 2.6.2.2. 'Smart' SMART

One advantage of SMART can be described as its ability to 'normalise' trajectories i.e. trajectories are scaled to unite, which facilitates the comparison of different trajectories using various scales. Another words, if two trajectories A and B are homothetic and for instance A is scaled to a range of ordered values, it is expected that the maximum similarity be seen when A is scaled to 1 if both A and B had already scaled to maximum before test scaling. Otherwise, the maximum similarity would be at the inverse scale of A to B. Then it would be difficult e.g. to see whether the maximum similarity was the result of homothetic geometry or only a coincident.

### 2.6.3. PCA and Scaling Integration

PCA is very commonly used to summarise complex multidimensional data sets which are often seen during metabonomic studies. Consistent differences in the average position or size of distribution between two clusters within the data will bias PCA to describe these differences at the cost of accurately modelling the treatment-related effects (Keun, Ebbels et al. 2004). To investigate how SMART affect homothetic trajectories, first, two curves (signals) with two different magnitudes were generated using Equation 4. Next, two additional curves were generated by multiplying the original curves by 2. Therefore, the trajectory of the second set of curves had twice magnitude of the trajectory of the original set. Then, a number of realisations were added to the data points of the smaller and the larger trajectories shown in Figure 14 A and B, respectively, using Equation 5. Subsequently, mean trajectories of the two data sets (the larger and the smaller) were calculated (Figure 14, E and F respectively). The first time point of each of these two mean trajectories as well as the original trajectory (without realisations and enlargement) were subtracted from the rest of their time points (translated to the first time point) and then they were scaled to maximum (Figure 14, G).

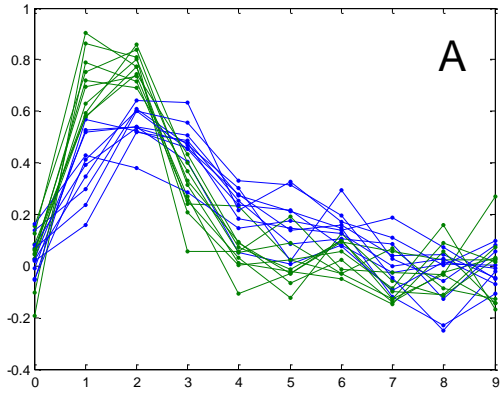
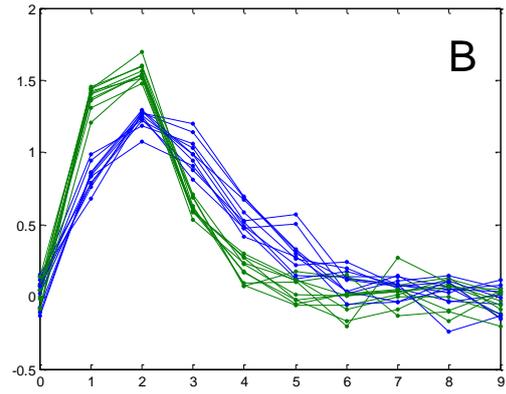
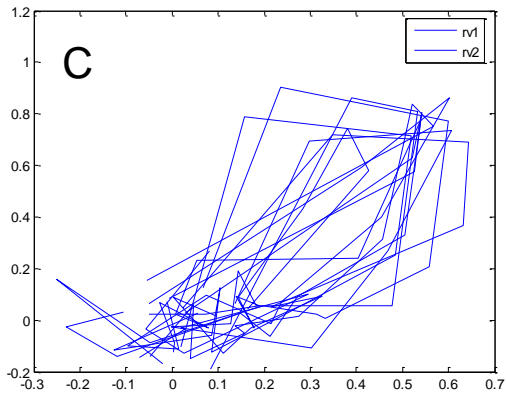
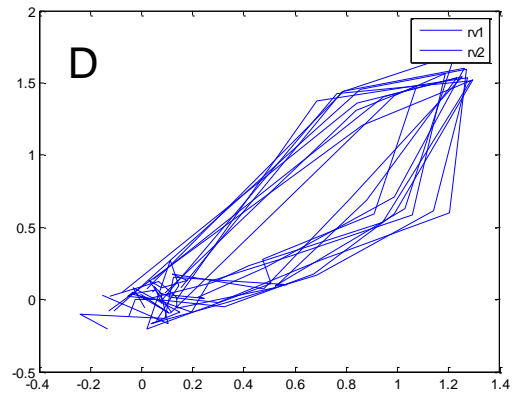
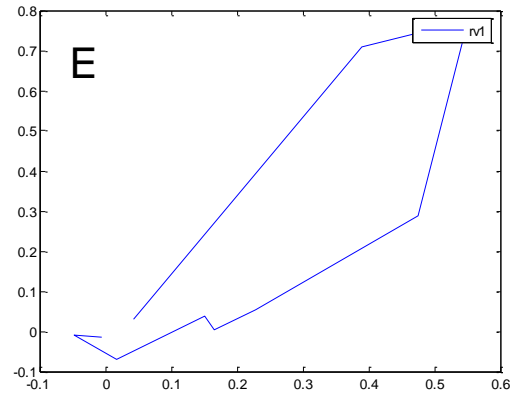
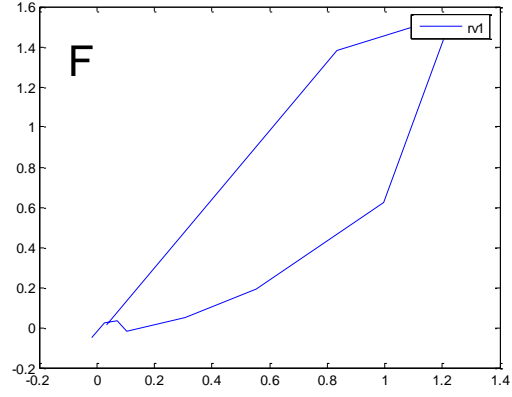

**Red Trajectory: SMART of *means* of smaller trajectory**
**Green Trajectory: SMART of *means* of bigger trajectory**
**Blue Trajectory: SMART of *original* trajectory i.e. *before adding noises***

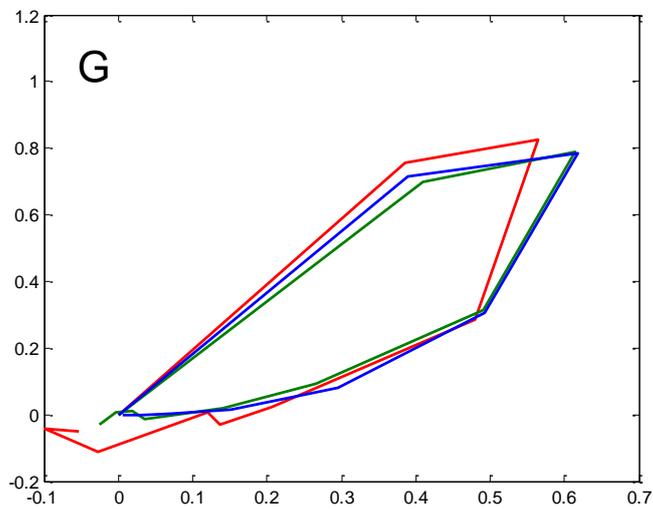

**Figure 14 Investigation of SMART technique**

To examine the collective effect of PCA and SMART, initially, two data sets were generated modelling two groups of animals. SMART scaling was carried out before and after PCA as shown in Figure 15 A and B, respectively illustrating 'first SMART then PCA' and 'first PCA then SMART' respectively.

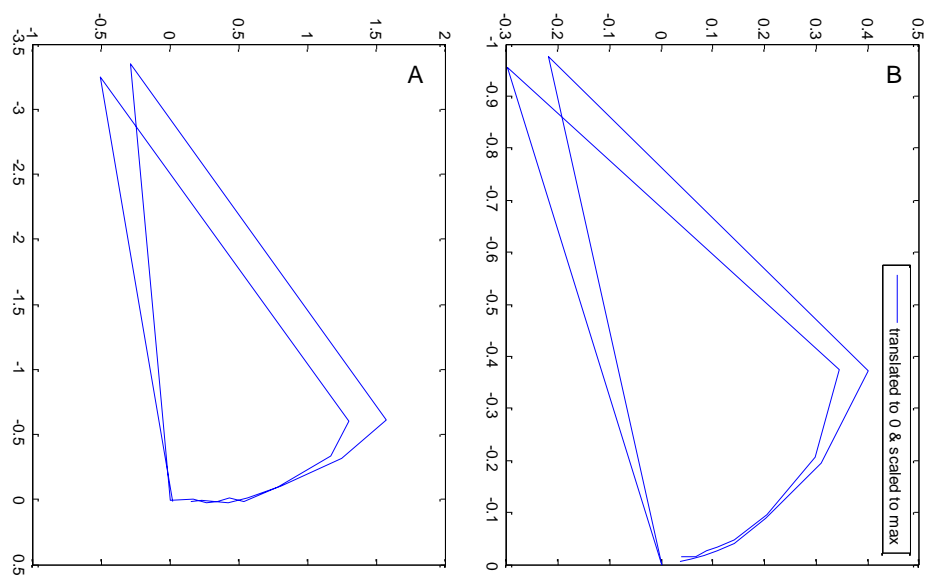

**Figure 15: PCA before and after SMART**

### *2.6.4. Homothetic geometry & metabolite speed*

Figure 16 A, B, C, and D, schematically illustrate trajectory generation from simulation of metabolic path, sampled signals in time series, and trajectories before and after SMART scaling, respectively. Modelled trajectories only differ in their initial speed and although the original dimensions follow exactly the identical path (Green line in A), the generated trajectories are similar (Weisstein; Johnson 1929) and do not have parallel sides (similitude center instead of homothetic centre) as shown in Figure 16 C and D. Thus, the time gap alone can lead to bifurcation of trajectories. SMART scaling does not rectify the problem. This is conceptually associated to the translation models presented earlier.

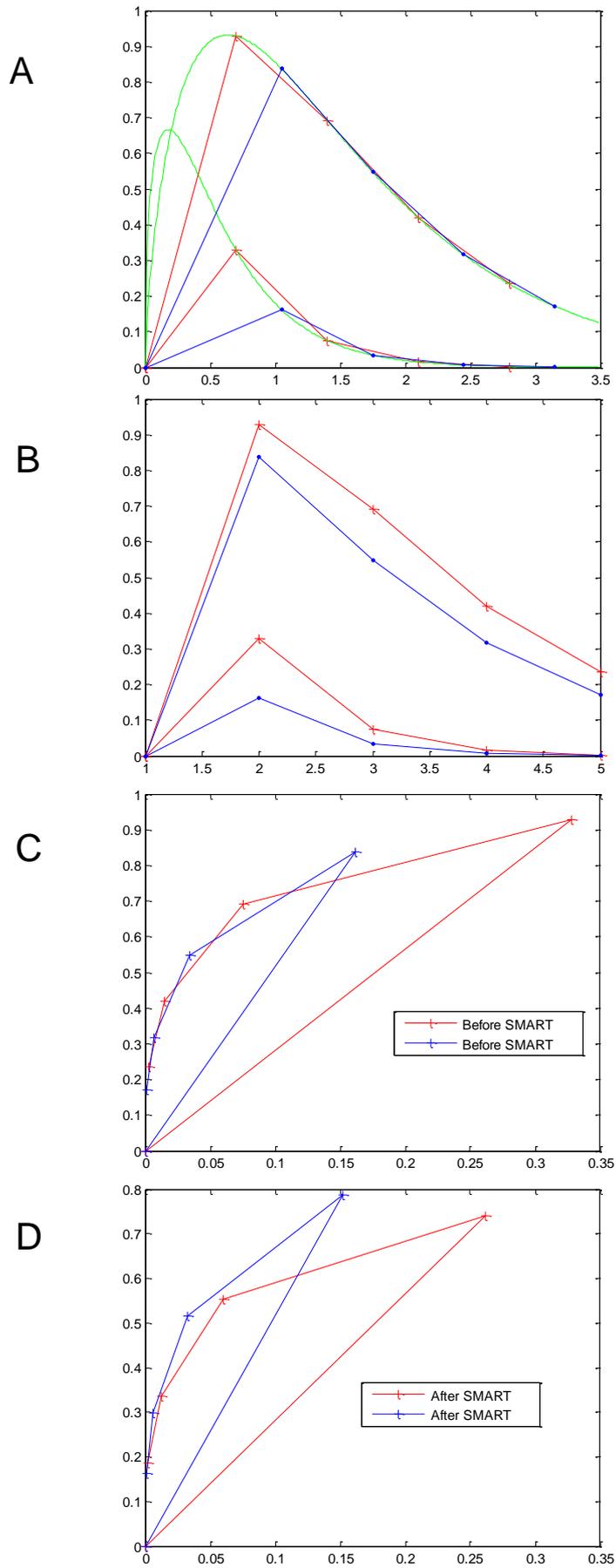

**Figure 16:Speed difference, homothetic geometry, and SMART**

## *2.7.  Simulation*

Using developed programmes, a number of sets of data were generated so as to test the potential effects of different factors including the level of noise, variations between animals, and variations between groups of test and reference animals (assuming they are from different species). In the following examples, data initially generated at 210 dimensions, of which each set of 7 dimensions were highly correlated (generated using Equation 5 ($i = 7$) for 30 original dimensions generated using Equation 4) such a way that when the level of noise is set to zero, those groups of 7 dimensions correlated 100% and the more the level of noise increases, the correlation coefficient among them declines. Structurally, data was divided into two groups (i.e. Test and Ref) which were identical unless the 'variation between groups' is set above zero. Next, PCA was performed on the generated data with two principle components. Subsequently, the data set (modelling test group) multiplied to an ordered range of scalars and each time the similarity was measured using CLOUDS-Overlap, SEA-Local, and SEA-Global. Figure 13-19 plot the influence of level of noise, variations between animals, variations between Ref and Test groups, and initial speed differences on similarity measures. The relative values of these factors displayed on the figures. In each figure, a and b illustrate translations to the first time point and to median respectively. Test trajectories are shown in blue colour and Ref trajectories in red. C, d, and e depict plots of similarity measures calculated based on CLOUDS-Overlap, SEA-Local, and SEA-Global respectively against the ranges of scales of homothetic figures of test trajectories. When, variations are close to zero, desirable results for CLOUDS and SEA (Local/Global) are 1 and 0 respectively.

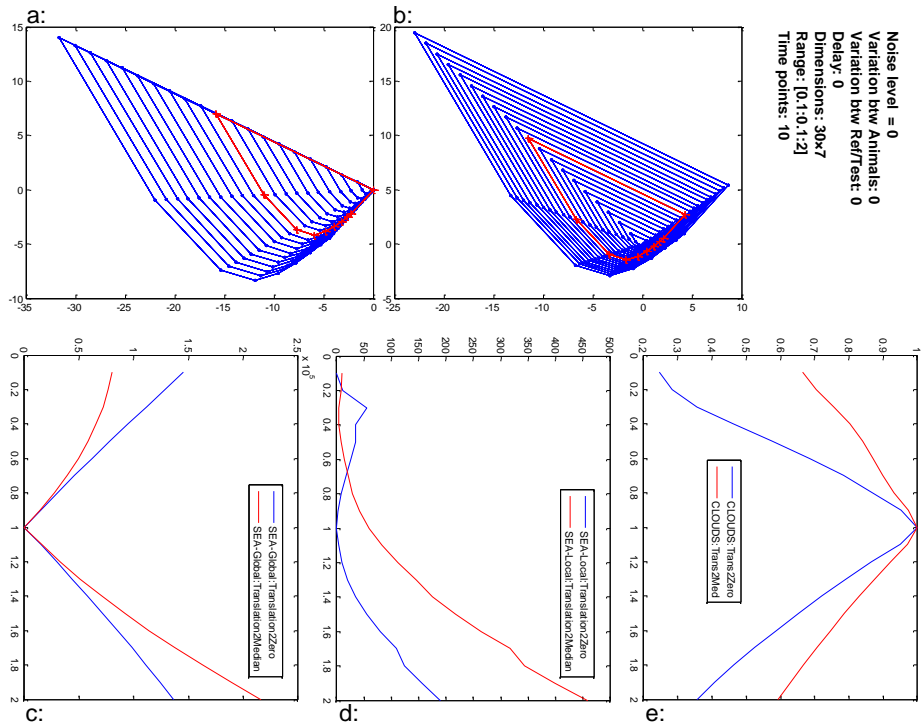

**Figure 17**

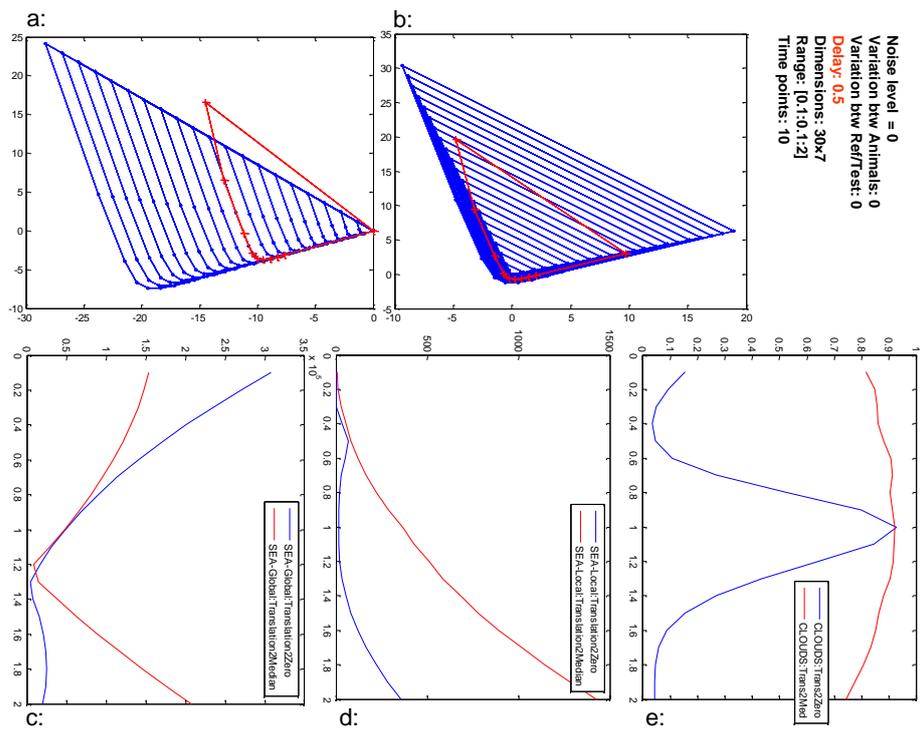

**Figure 18**

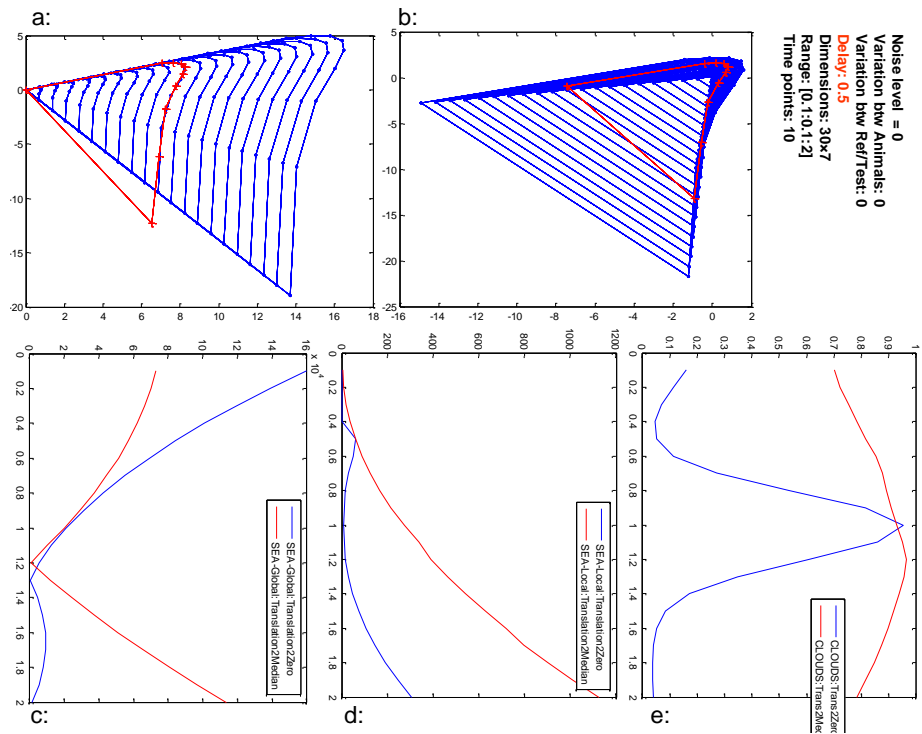

**Figure 19**

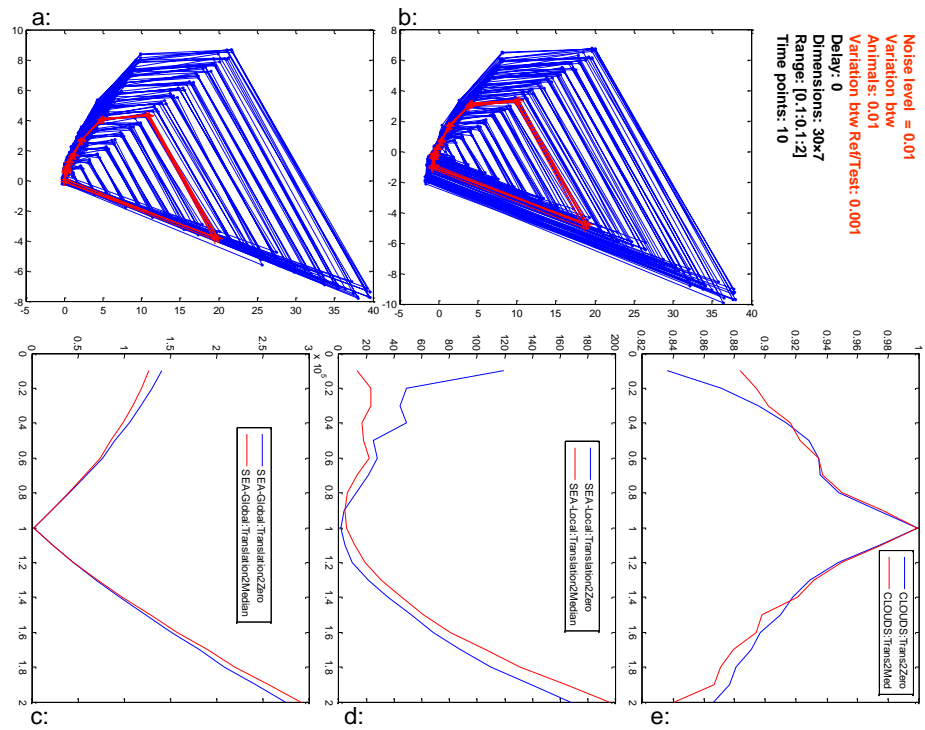

**Figure 20**

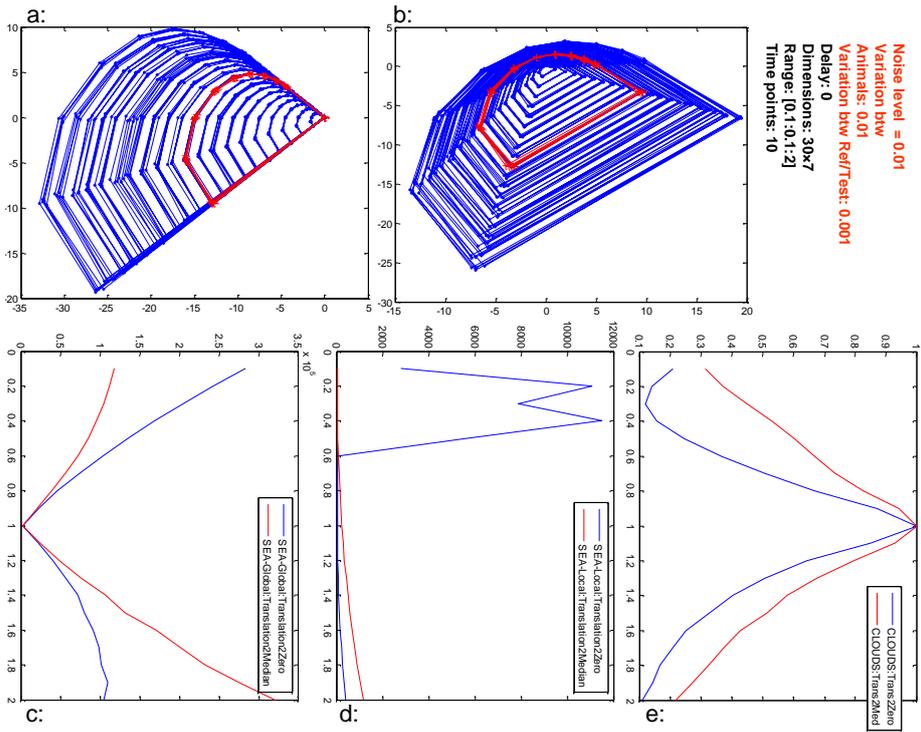

**Figure 21**

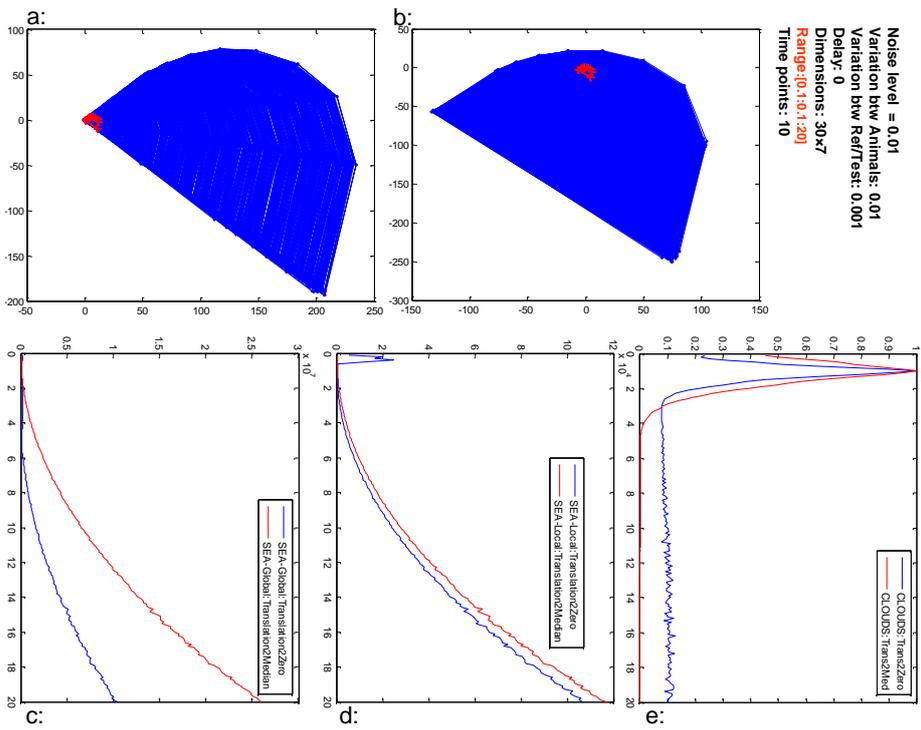

**Figure 22**

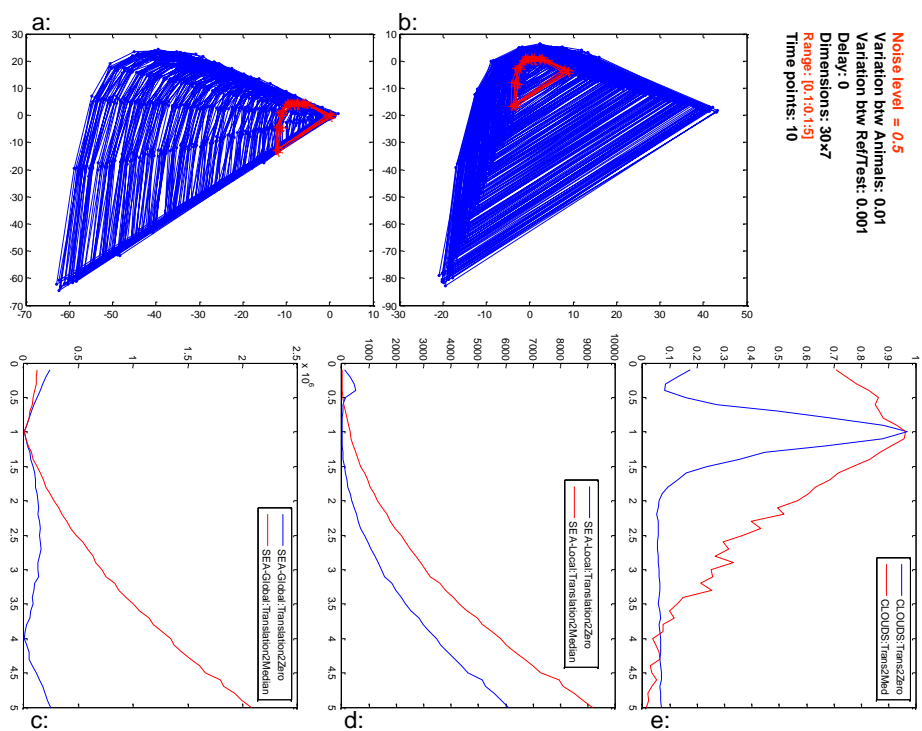

**Figure 23**

## 2.8. Experimental (biological) data:

Urine samples NMR collected from 5 experiments studying metabolic response of hydrazine carried out by 5 participating companies participating in the COMET project, (Lindon, Keun et al. 2005) were included in analysis. The studies labelled as L01, N01, R01, R02, S01, and R12. For each experiment, 20 rats were randomly chosen and divided into control and dosed groups and their urine samples were collected at 10 time points during 8 days. Two samples were taken before administrating of Hydrazine. Half of the animals receiving hydrazine were euthanized at 48h post-dose and the other half euthanized at 168h post-dose for histopathological and clinical chemistry evaluation.

1H NMR spectra were measured at 600 MHz and 300K using a robotic flow-injection system (Bruker Biospin, Karlsruhe, Germany). Signals in regions of equal width (0.04 ppm) in the chemical shift ranges 0.20-4.50ppm and 5.98-10.0ppm were integrated and spectral artefacts in the region 4.50-5.98ppm were excluded and finally after replacing non-endogenous signals related to drug related compounds (Ebbels 2002) (thus resulting analysis is focused on reliably profiling metabolic change within the living system), each NMR spectrum was reduced into 190 variables. All reduced spectra were then normalized to a constant integrated intensity of 100 units to take account of large variations in urine concentration. PCA carried out on all data at

# 3. Results & Discussion:

A pervasive theme in metabolic trajectory analysis is judging when two or more responses are the same as it has been hypothesized that the geometry of the metabolic trajectories is characteristic of the biological response (Keun, Ebbels et al. 2004). Yet, due to unavoidable inter-individual variations, the exact trajectories characterising the biological responses differ. Therefore, it is intriguing to examine whether the minor differences seen between metabolic trajectories of a specific treatment, correspond to the variations seen in other biological manifestations of the same treatment. Differences in trajectories can be measured via alignment procedures introduced and implemented in this study.

Trajectory alignment yields 2 numbers which can be termed as 'best scale' and 'maximum similarity measure'. Scales and measures were obtained after PCA using 2 and 5 PCs. The hypothesis that relative scales and/or similarity of a metabolic trajectories are indicative of variations in underlying biological response, the best scales and the maximum similarity measures from the CLOUDS and the $SEA_{Global}$ as well as largest magnitudes from SMART method were compared with the severities of the two types histopathological lesions (i.e. lipidosis and glycogen deposition) in hepatic cells observed 48 and 168 hours after administration of Hydrazine to the experimental rats in high dose group. Significant correlation was observed between the severity of lipidosis and glycogen lesions at 48 hours and scales in both 2PCs and 5PCs analysis. Severity of lesion was rated using the following scale: 0 – not observable; 1 – minimal; 2 – mild; 3 – moderate; 4 – marked; 5 - severe. The $SEA_{Global}$ values (SSE) showed significant correlation with glycogen deposition at the same hour. No significant correlation was observed with the rest of variables. Table 1 summarises the results of the correlation analysis. Table 1 presents correlation matrix between histopathological lesions only for comparison purpose, which shows significant correlation between the two types of lesions at 48 hours. Figure 24, Figure 25, Figure 26, and Figure 27 respectively scatter plot the 3 three observed significant observations as well as scatter plot of the CLOUDS similarities and the glycogen deposition at 48 hours (in which a significant correlation exists between scales derived from the CLOUDS method the lesion). All figures contain only 5 points. In Figure 24, Figure 25, and Figure 26, floor effect is visible because in glycogen deposition data the real difference is between R01/R02 and the rest of values gives correlation. Figure 27 related to lipidosis demonstrate an obvious correlation between the severity of the lesions and the scales of the trajectories obtained via the CLOUDS method..

**Table 1: Correlation matrix between the severity histopathologic lesions and the similarity measures**

| Five Principle Components | SEA Scales | SEA Values | CLOUDS Scales | CLOUDS Values | SMART Largest Magnitude |
|---|---|---|---|---|---|
| **Lipidosis HD 48h** | -0.37635 | 0.27419 | -0.82568 | 0.7451 | 0.30691 |
| *p. value* | *0.53238* | *0.65532* | *0.085044* | *0.14844* | *0.61546* |
| **Lipidosis HD 168h** | -0.42034 | 0.61414 | -0.623 | 0.55692 | -0.24263 |
| *p. value* | *0.48102* | *0.27045* | *0.26159* | *0.32949* | *0.69414* |
| **Glycogen Deposition HD 48h** | 0.37998 | -0.3713 | **0.96844** | -0.75015 | -0.43828 |
| *p. value* | *0.52811* | *0.53835* | ***0.0066996*** | *0.14416* | *0.46039* |
| **Glycogen Deposition HD 168h** | 0.0055559 | -0.44825 | 0.629 | -0.37613 | -0.35705 |
| *p. value* | *0.99293* | *0.44901* | *0.25563* | *0.53264* | *0.55524* |
| **Two Principle Components** | **SEA Scales** | **SEA Values** | **CLOUDS Scales** | **CLOUDS Values** | **SMART Largest Magnitude** |
| **Lipidosis HD 48h** | -0.44387 | 0.726 | -0.80552 | -0.18649 | 0.25827 |
| *p. value* | *0.454* | *0.16492* | *0.099894* | *0.76394* | *0.67486* |
| **Lipidosis HD 168h** | -0.55161 | 0.65903 | -0.52198 | 0.097037 | -0.2715 |
| *p. value* | *0.33512* | *0.22638* | *0.36695* | *0.87664* | *0.65861* |
| **Glycogen Deposition HD 48h** | 0.52546 | **-0.86745** | **0.97435** | 0.44567 | -0.32047 |
| *p. value* | *0.36318* | ***0.056762*** | ***0.0049134*** | *0.45194* | *0.59906* |
| **Glycogen Deposition HD 168h** | 0.56627 | -0.68447 | 0.67076 | 0.77774 | -0.1709 |
| *p. value* | *0.31964* | *0.20239* | *0.21523* | *0.1215* | *0.78347* |

**Table 2: Correlation matrix between histopathology findings lesions**

| | Lipidosis HD 48h | Lipidosis HD 168h | Glycogen Deposition HD 48h | Glycogen Deposition HD 168h |
|---|---|---|---|---|
| **Lipidosis HD 48h** | 1 | 0.79642 | **-0.9118** | -0.10693 |
| *p. value* | *1* | *0.10683* | ***0.031023*** | *0.86411* |
| **Lipidosis HD 168h** | 0.79642 | 1 | -0.61237 | 0.069393 |
| *p. value* | *0.10683* | *1* | *0.27223* | *0.91172* |
| **Glycogen Deposition HD 48h** | **-0.9118** | -0.61237 | 1 | 0.48869 |
| *p. value* | ***0.031023*** | *0.27223* | *1* | *0.40352* |
| **Glycogen Deposition HD 168h** | -0.10693 | 0.069393 | 0.48869 | 1 |
| *p. value* | *0.86411* | *0.91172* | *0.40352* | *1* |

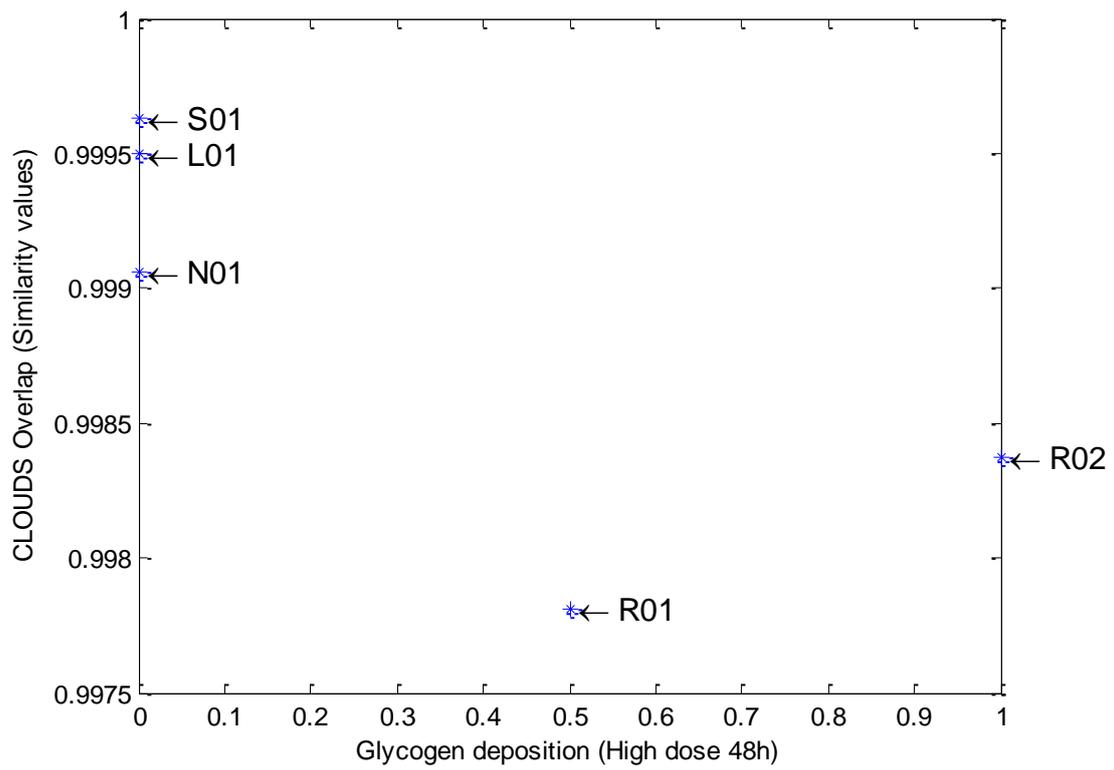

**Figure 24**

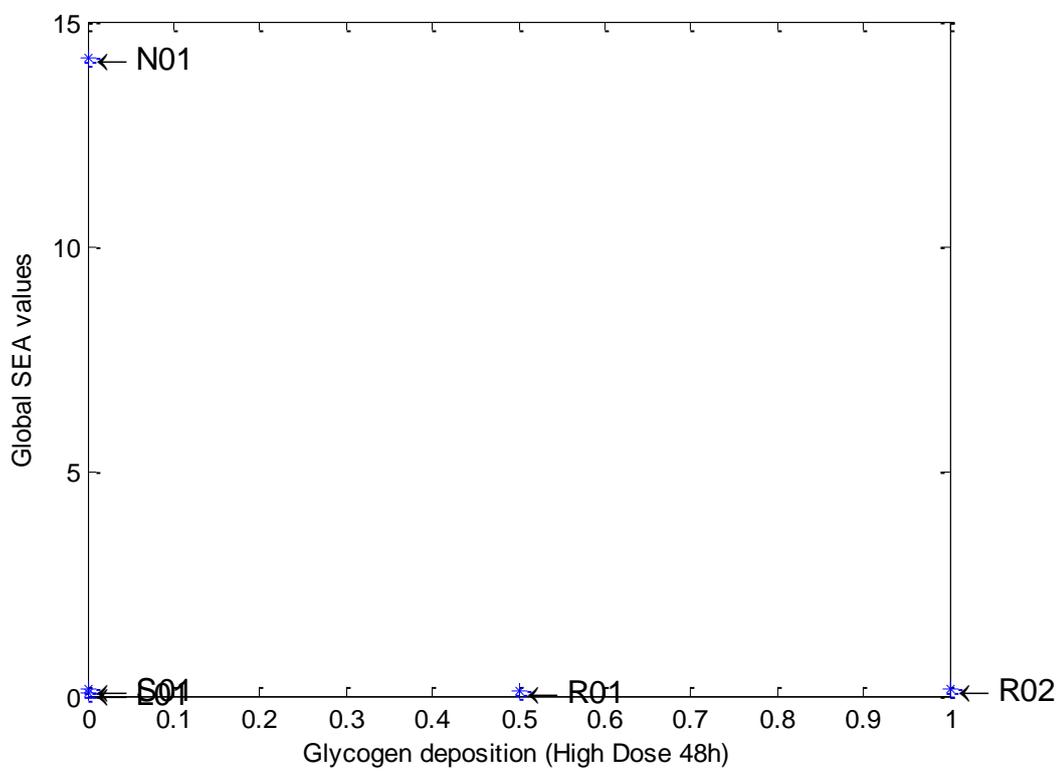

**Figure 25**

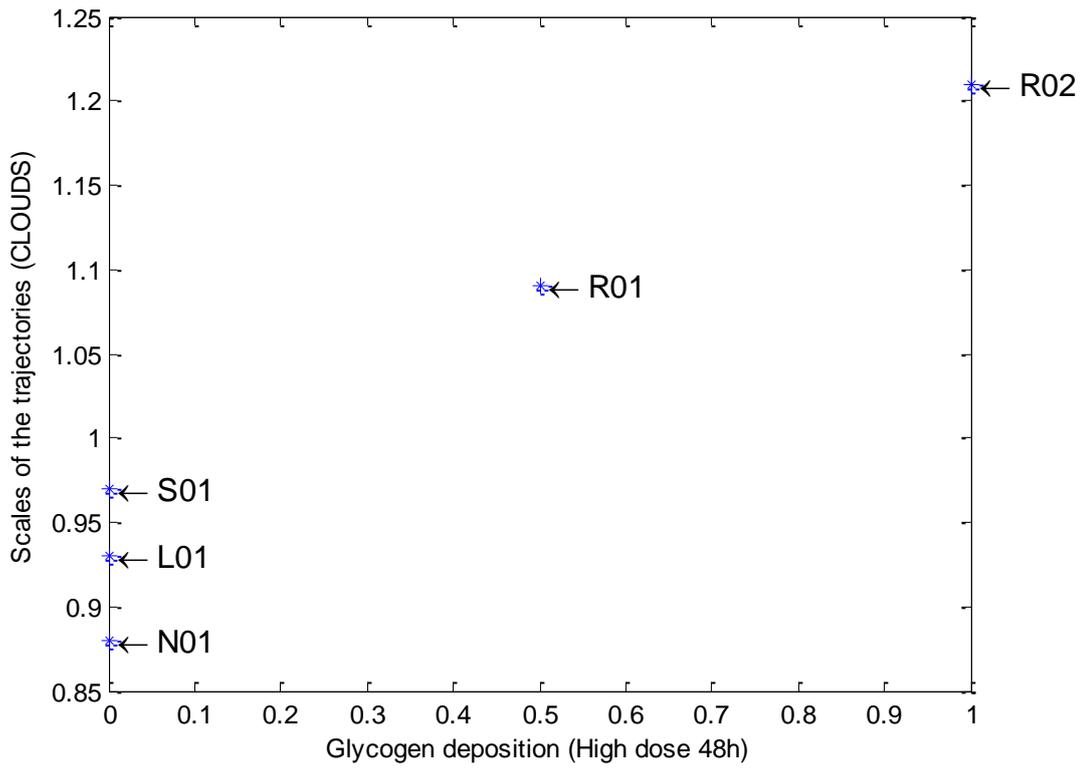

**Figure 26**

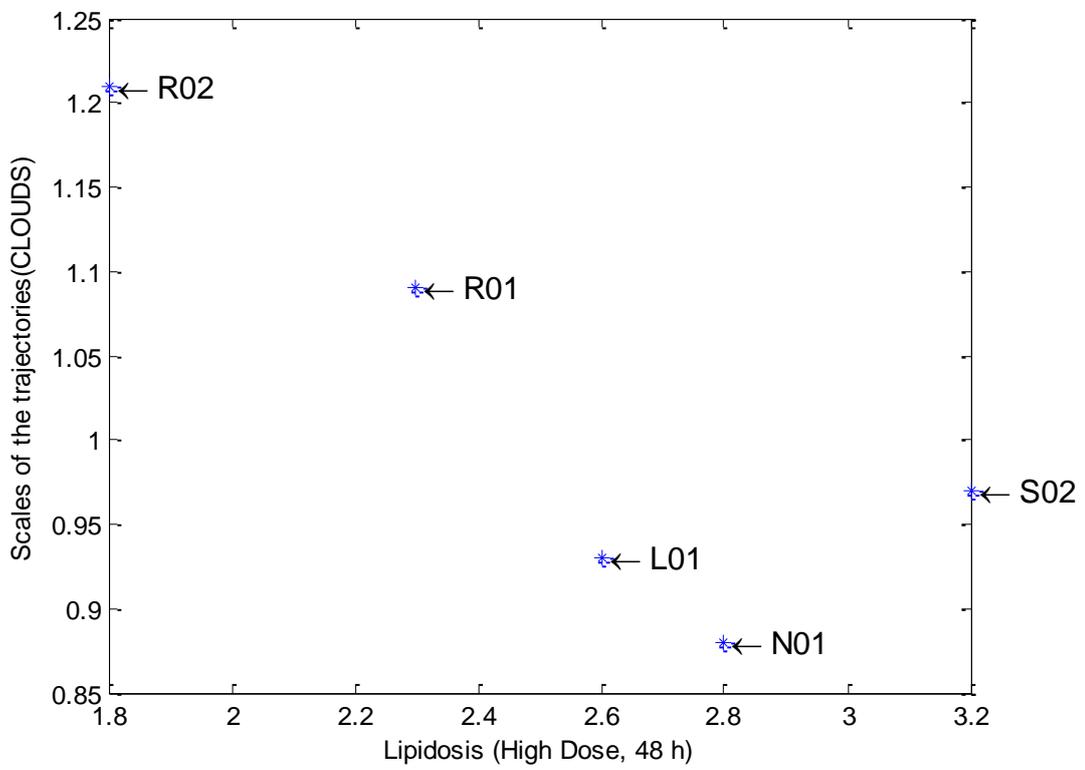

**Figure 27**

## 3.1. Homothetic geometry & SMART & PCA

Although conceptually the drawing conclusion is true that in principal homothetic geometry is independent of sampling and principle and in principle homothetic similar trajectories,

In principle, rotationally transformed similar trajectories would not share the same directions of metabolite changes and thus would not be expected to reflect similar system behaviour (Keun, Ebbels et al. 2004) and also homothetic geometry is independent of sampling. However, during interpretation of metabolic trajectory, it is important to bear in mind that in 'real world application', unavoidable onset/speed variations in biological responses from of an identical system behaviour together with sampling effects are added complexity which can make the trajectories appear non-homothetic as shown in Figure 28 and Figure 29 two trajectories of identical system behaviour with mild and extreme divergences respectively. The largest magnitude of identical response can be significantly distorted i.e. they are 2.61 and 2.18 for the Figure 28 and 1.35 and 0.36 for Figure 29 in the respectively. This means the largest magnitude of in one of the slower model has shown 1.1972 and 3.7500 times (almost 4 fold increase) bigger whereas when the SMART method is used to measure similarities between trajectories it would have been expected they show almost identical magnitudes.

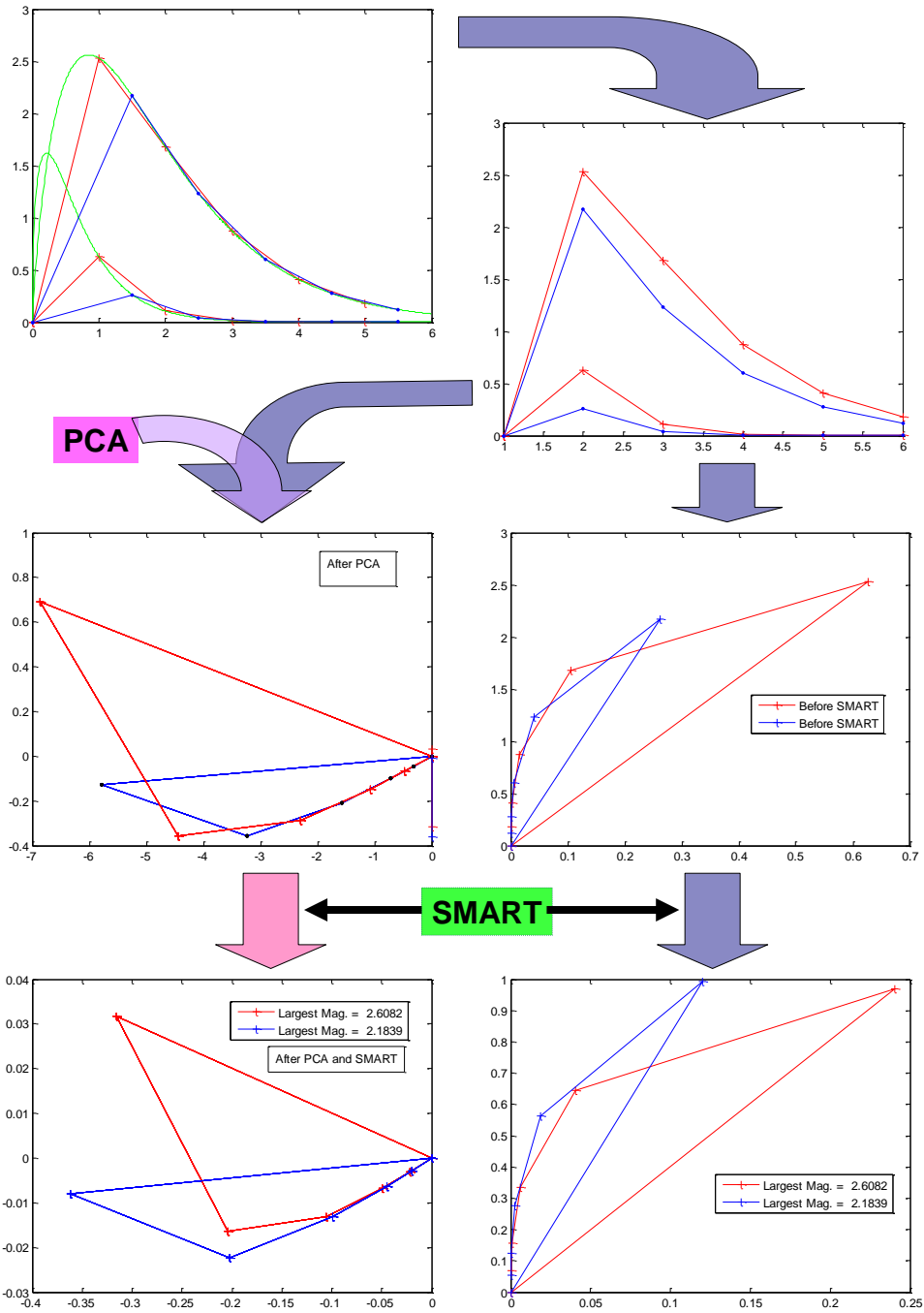

**Figure 28: Simulation of two trajectories of identical system behaviour with variation in speed of response before and after PCA**

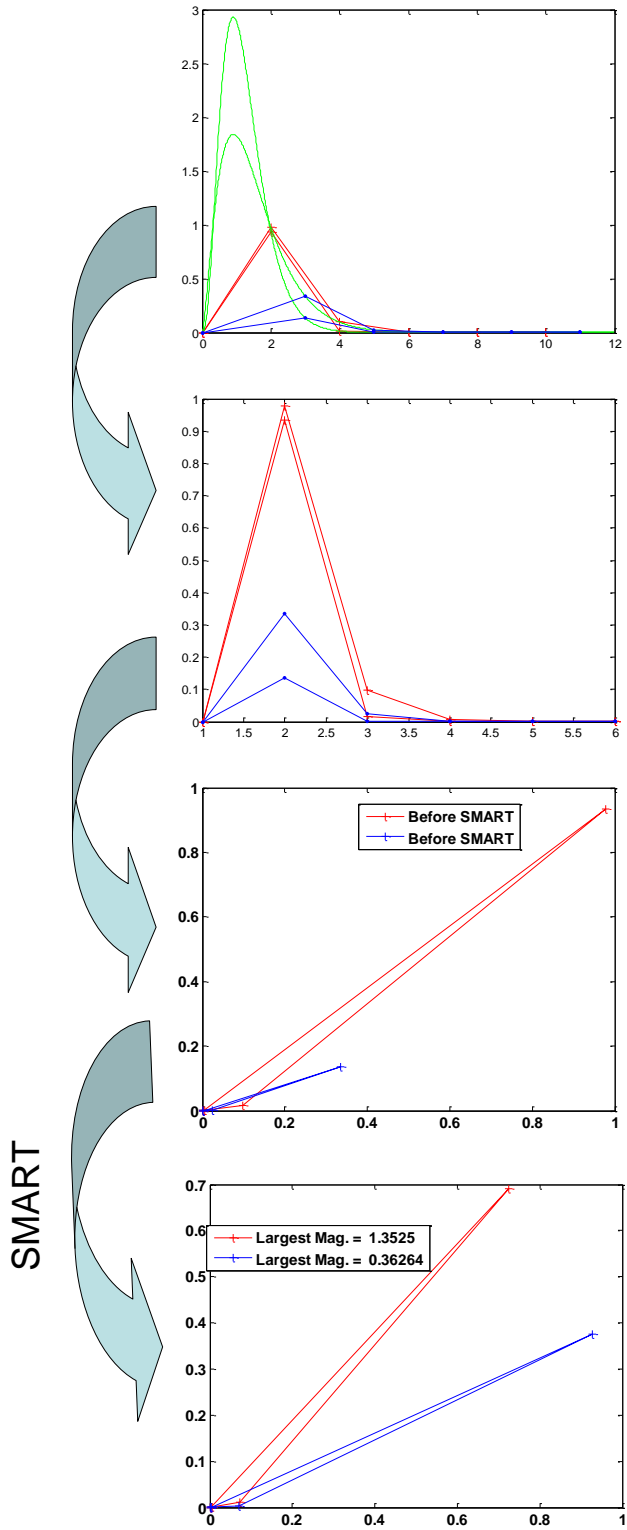

**Figure 29: two trajectories of identical system behaviour with extreme divergence**

In conclusion; the results reinforce the hypothesis that differences between metabolic trajectories rises from inter-individual variations in biological responses to a specific treatment which to reasonable extents can be eliminated by alignment procedure presented in this study especially the CLOUDS method. Aligned trajectories, characteristic of a specific treatment, then can be used for comparison with other treatment specific or unknown trajectories, which can have many metabonomic applications such as preclinical toxicological screening. These results further confirm that concept of 'likeness' in metabonomic trajectories is biologically relevant.